\documentclass[aps,pra,reprint,amsmath,amssymb,longbibliography,floatfix]{revtex4-2}
\usepackage[T1]{fontenc}
\usepackage[utf8]{inputenc}
\usepackage{bm}
\usepackage{amsthm}
\usepackage{booktabs}
\usepackage{graphicx}
\usepackage{tikz}
\usepackage{pgfplots}
\usetikzlibrary{arrows.meta,positioning}
\pgfplotsset{compat=1.18}
\usepackage[hidelinks]{hyperref}
\newcommand{\Ftwo}{\mathbb F_2}
\newcommand{\SO}{\operatorname{SO}}
\newcommand{\GL}{\operatorname{GL}}

\newcommand{\Tr}{\operatorname{Tr}}

\newcommand{\ket}[1]{|#1\rangle}
\newcommand{\bra}[1]{\langle#1|}
\newcommand{\ip}[2]{\langle #1,#2\rangle}
\newcommand{\aEight}{\ket{a_8}}
\newcommand{\eps}{\varepsilon}
\newcommand{\cG}{\mathcal G_{\mathrm{conv}}}
\newcommand{\Rthree}{\mathcal R_3}
\newcommand{\Rfour}{\mathcal R_4}
\newtheorem{theorem}{Theorem}

\newtheorem{corollary}[theorem]{Corollary}

\begin{document}
\title{Reducing the cost of ancilla purification for Majorana quantum computation}

\author{Piotr {\'C}wikli{\'n}ski}
\affiliation{International Centre for Theory of Quantum Technologies, University of Gda{\'n}sk, ul. prof. Marii Janion 7, 80-309 Gda\'nsk, Poland}

\author{\L{}ukasz Fojcik}
\affiliation{Computational Geology Laboratory, Polish Geological Institute, al. Jaworowa 19, 53-122 Wroc\l{}aw, Poland}

\author{Micha{\l} Studzi{\'n}ski}
\affiliation{International Centre for Theory of Quantum Technologies, University of Gda{\'n}sk, ul. prof. Marii Janion 7, 80-309 Gda\'nsk, Poland}

\date{\today}

\begin{abstract}

Braiding Majorana zero modes does not by itself provide a universal set of
quantum gates. Additional operations can be supplied through ancillary
states, whose preparation need not have the same protection as braiding.
Here, we consider purification of the non-Gaussian ancilla used for an
entangling gate in Bravyi's scheme. We keep his elementary braid-and-fusion
comparison and change how its outputs are combined. An additional layer
compares the product of the three stabilizer signs. The first accepted
errors giving an incorrect output then require three faulty input blocks
rather than two. As a result, the preparation error decreases cubically
instead of quadratically, and the asymptotic raw-copy overhead is reduced.
We also construct a preliminary sequence in which a processed block is
paired with a fresh copy from the same noisy source. For independent inputs supported in the even-parity sector and ideal
purification operations, this sequence establishes a sufficient
distillability fidelity of 59.95 percent after free Gaussian twirling. At a common source fidelity, it reduces the expected cost
of preparing an input for the recurrence by more than seven orders of
magnitude relative to the balanced construction studied here. Finally,
we examine imperfect purification operations in an effective syndrome-noise
model. The additional layer leaves the leading operation-error coefficient
unchanged, although a nonzero error floor remains.

\end{abstract}
\maketitle

\section{Introduction}\label{sec:intro}

Topological quantum computation uses the exchange of non-Abelian anyons
to act on encoded quantum information. In Majorana-based schemes,
braiding realizes Ising-anyon statistics and provides a protected but
restricted set of gates
\cite{kitaev2001,ivanov2001,nayak2008,alicea2012,beenakker2013}.
These gates are not sufficient for universal computation. The missing
operations must therefore be supplied in another way. One possibility
is to prepare additional ancillary states using unprotected operations
and then purify them using braiding and fusion measurements
\cite{bravyi2006}. The accuracy and cost of ancilla preparation then
become part of the requirements for computation.

The state considered here is Bravyi's eight-Majorana, or four-mode,
ancilla. In his encoding, a purified copy allows an entangling Clifford
gate to be implemented; a separate phase ancilla is required for the
non-Clifford part of the computation \cite{bravyi2006}.
We study the preparation of the first of these resources. There are two
questions: how noisy can the initial ancillas be and still admit
purification, and how many copies are needed to reach a given output
accuracy? A protocol may answer the first question favorably and still
be expensive to use.

The allowed operations are important for both questions. Quadratic
Majorana Hamiltonians generate fermionic linear optics (FLO), and
circuits with Gaussian inputs and FLO operations admit efficient
classical descriptions
\cite{terhal2002,knill2001,valiant2002,bravyi2005,jozsa2008}.
The relation to universal computation depends on which additional gates,
states, and geometries are available, as in matchgate computation
\cite{bravyi2002,brod2011,brod2014}.
In the setting studied here, the supplied ancilla is non-Gaussian, while
its purification uses Gaussian operations. Bravyi's construction moves
the operation unavailable within FLO to the preparation of a state;
the subsequent comparisons use braids, bilinear fusion measurements,
classical feed-forward, and postselection \cite{bravyi2006}.

There is a corresponding restriction on the input states. Convex mixtures
of Gaussian fermionic states remain classically simulable under FLO
\cite{demelo2013}, and their boundary can be characterized in the
four-mode case by a concurrence criterion \cite{oszmaniec2014}.
Every pure non-Gaussian fermionic state is a magic state for matchgate
computation \cite{hebenstreit2019}, but the mixed-state question is
different: being outside the convex-Gaussian set does not by itself
establish distillability. Gaussian decompositions and simulation methods
with non-Gaussian inputs give further ways to study this distinction
\cite{cudby2023,dias2024,reardon2024}; non-Gaussian inputs also serve as
resources in fermion sampling \cite{oszmaniec2022}.
Related questions arise in bosonic Gaussian no-distillation theorems
\cite{eisert2002,giedke2002} and in qubit magic-state resource theories
\cite{veitch2014,howard2017,campbell2010}, although their conclusions
do not directly settle the fermionic purification problem considered here.

Here, we keep Bravyi's elementary two-copy comparison and change the way
in which it is used. In the even-parity sector, the signs of three
commuting stabilizers label the possible errors of an ancilla. The
comparison tests whether one chosen stabilizer product has the same sign
on two input blocks. A disagreement is rejected; on agreement, one block
is retained with a known Gaussian correction. Changes of Majorana frame
allow the same circuit to compare different stabilizer products.
Bravyi's three-layer round already retains the intermediate syndrome
distribution and randomizes it only after the complete round
\cite{bravyi2006}. We follow the same convention.

Our first modification is to add a fourth comparison layer. It acts on
two independently prepared outputs of the first three layers and compares
the product of all three stabilizers. This product contains no independent
syndrome bit, but the comparison detects wrong outputs that survive the
earlier layers. In the four-layer tree, every pattern with two faulty raw blocks
either is rejected or gives the correct final syndrome. An incorrect accepted output first appears with three faulty
blocks. The preparation error therefore decreases cubically rather than
quadratically. The same modification lowers the unstable fixed point
from about 61.56 percent to about 60.47 percent and reduces the asymptotic
number of copies needed for high-accuracy purification.

Below this fixed point, a different initial sequence is needed before
the recurrence becomes useful. We give a balanced twenty-five-layer
construction which crosses the boundary at a raw fidelity of 59.97
percent, but has a very large expected copy cost. We then allow one
stage to pair a processed block with a fresh raw block, rather than with
a second copy of the processed state. An explicit twelve-stage sequence
works at 59.95 percent. At the common raw fidelity of 59.97 percent,
its expected cost of preparing an input for the recurrence is smaller
by more than seven orders of magnitude. This comparison is between the
two stated sequences, not between optimal protocols of the two types.
Both use the same source and the same elementary comparison.

Separating this initial preparation from the later high-accuracy rounds
is useful, as it is in entanglement purification
\cite{bennett1996a,bennett1996b,deutsch1996} and magic-state distillation
\cite{bravyikitaev2005,reichardt2005,bravyihaah2012,meier2013}.
Pairing a processed resource with a fresh auxiliary input also has a
precedent in entanglement purification \cite{dur1999}.
Here, the improvement depends on the fermionic syndrome dynamics and
on retaining unequal error populations between comparisons. The
comparison order matters: the fidelity can decrease at an intermediate
stage even when the complete sequence prepares a state that can be
purified. We compare the full protocols at fixed source fidelity and
output accuracy, as well as their asymptotic error laws.

We also consider imperfect purification operations. In an effective
model with a syndrome error after each accepted comparison, both the
three- and four-layer rounds have the same leading operation-error
coefficient. The extra layer improves suppression of preparation errors
without removing the error floor from operation faults. The
heterogeneous initial sequence is substantially less sensitive than the
long balanced sequence at the same raw fidelity. These results concern
an ancilla-preparation model rather than fault tolerance of a complete
device. Proposals based on superconducting islands, Majorana boxes,
and hybrid topological--stabilizer schemes
\cite{aasen2016,karzig2017,plugge2017,litinski2017}, Majorana codes and
surface-code constructions \cite{bravyicodes2010,vijay2015}, and
circuit-level noise analyses \cite{li2016} address additional requirements
of such a device.

The Gaussian frame changes used here are already available in Bravyi's
scheme. We describe their action through affine transformations of the
syndrome labels; the associated finite geometry is related to the
standard $\mathrm{RM}(1,3)$ code \cite{macwilliams1977}.
The new ingredients are the four-layer recurrence and the explicit
heterogeneous initial sequence, together with their error and cost
analysis. For the input family studied here, the convex-Gaussian boundary
lies at fidelity one half \cite{oszmaniec2014}. Our sufficient fidelity
of 0.5995 also applies to arbitrary independent even-parity inputs
having this overlap with the target, by Gaussian twirling. It leaves
a region above the convex-Gaussian boundary in which distillability
remains unresolved.

The paper is organized as follows. Section~\ref{sec:model} introduces the
input state and the physical two-copy comparison.
Section~\ref{sec:checks} explains how Gaussian frame changes select the
stabilizer product to be compared. The four-layer recurrence and the
origin of its cubic error suppression are presented in
Sec.~\ref{sec:recurrence}. Section~\ref{sec:certificate} gives the
balanced and heterogeneous initial sequences.
Section~\ref{sec:cost} compares their resource costs and discusses the
use of the purified state. Imperfect operations are considered in
Sec.~\ref{sec:noise}, followed by the conclusions in
Sec.~\ref{sec:discussion}. The derivations and numerical details are
collected in the Appendices.

\section{Physical setting and the elementary purification step}\label{sec:model}

Let $c_1,\ldots,c_8$ be Majorana operators,
$c_p^\dagger=c_p$ and $\{c_p,c_q\}=2\delta_{pq}I$, describing four
fermionic modes. One such group of eight Majoranas will be called a
block. Besides the noisy resource states, we allow fermionic Gaussian
unitaries and ancillas, bilinear projective measurements, classical
feed-forward, postselection, and discarding of modes. These are the FLO
operations used in Refs.~\cite{bravyi2005,demelo2013}.

The resource state is Bravyi's four-mode ancilla. In our convention it is
the common positive-eigenvalue state of
\begin{align}
 S_1&=-c_1c_2c_5c_6,& S_2&=-c_2c_3c_6c_7,\nonumber\\
 S_3&=-c_1c_2c_3c_4,& Q&=c_1c_2c_3c_4c_5c_6c_7c_8 .
 \label{eq:stabilizers}
\end{align}
Hence
\begin{equation}
 \Pi_0=\aEight\bra{a_8}
 =\frac1{16}(I+Q)\prod_{j=1}^3(I+S_j),
 \label{eq:target}
\end{equation}
and, in the standard occupation convention,
$\aEight=(\ket{0000}+\ket{1111})/\sqrt2$.

\subsection{Input noise model}

We work in the even-parity block. Denoting $I_+=(I+Q)/2$, the input states
are
\begin{equation}
 \sigma_F=F\Pi_0+\frac{1-F}{7}(I_+-\Pi_0),\qquad 0\leq F\leq1.
 \label{eq:modelb}
\end{equation}
The parameter $F$ is the overlap with the target, while the other seven
even-parity states have equal populations. We assume independent copies
of $\sigma_F$. The source may prepare the even sector directly, or it
may be selected by a nondestructive measurement of total fermion parity
before purification. The latter option requires an additional resource:
a nondestructive measurement of $Q$ is not a bilinear FLO measurement.

For comparison with a depolarized four-mode source, consider
$\rho_p=(1-p)\Pi_0+pI_{16}/16$. Conditioning on even parity gives
Eq.~\eqref{eq:modelb} with
\begin{equation}
 F_+(p)=\frac{16-15p}{16-8p},\qquad
 p(F)=\frac{16(1-F)}{15-8F}.
 \label{eq:depolarizing}
\end{equation}
This conversion includes parity postselection. The preparation bound
below applies to the conditioned even-sector state, not to purification
of the unconditioned $\rho_p$ using FLO alone.

\subsection{Three stabilizer signs as an error syndrome}

In the even block, the signs of $S_1,S_2,S_3$ label eight orthogonal
states. We write the signs as a three-bit syndrome
$s=(s_1,s_2,s_3)$, where $s_j=0$ corresponds to the positive eigenvalue
of $S_j$ and $s_j=1$ to the negative one. The target state has syndrome
$000$. For a syndrome-diagonal state,
$\rho_P=\sum_sP(s)\Pi_s$, while Eq.~\eqref{eq:modelb} corresponds to
\begin{equation}
 \pi_F(0)=F,\qquad
 \pi_F(s)=\frac{1-F}{7}\quad(s\ne0).
 \label{eq:isotropic}
\end{equation}
All additions of syndrome bits below are modulo two. These bits describe
errors within the fixed even-fermion-parity sector. Their sum is a
syndrome parity and must not be confused with the total fermion parity
$Q$, which has eigenvalue $+1$ throughout this sector.

Bravyi's elementary purification step acts on two four-mode blocks. It
compares one stabilizer sign and rejects when the signs disagree. If they
agree, one block is kept and a known Gaussian feed-forward correction is
applied. No quartic stabilizer is measured directly. The comparison is realized by a braid-and-fusion circuit with bilinear fusion
measurements. The corresponding physical identity is recalled in
Appendix~\ref{app:circuit}.

The syndrome description turns the calculation into error detection on
three binary labels. It does not assume that these labels are read out
on each block. The physical operation remains the braid-and-fusion
comparison, and the labels determine its acceptance and corrected output.

We call an input state postselection-distillable if, for every target
accuracy, a finite protocol using finitely many copies has a nonzero
probability of producing an output with that accuracy. This definition
does not require a positive asymptotic rate. Until
Sec.~\ref{sec:noise}, the purification operations are ideal and only the
input preparation is noisy.

\section{Changing the checked stabilizer with Gaussian operations}\label{sec:checks}

The original circuit compares one particular stabilizer sign. By changing
the Majorana frame before and after the circuit, the same physical
comparison can test other products of $S_1,S_2,S_3$. No new non-Gaussian
interaction is needed. This allows us to build the purification round
below from one elementary comparison.

\subsection{The base comparison}

Let the two input syndromes be $r$ and $s$. Bravyi's checksum accepts when
their third bits agree. After the outcome-dependent Gaussian correction,
the retained syndrome is
\begin{equation}
 (r_1+s_1,\ r_2+s_2,\ r_3).
 \label{eq:basecheck}
\end{equation}
The four bilinear fusion measurements, the checksum, and the conditional
double-braid corrections are listed in Appendix~\ref{app:circuit}. In
particular, Eq.~\eqref{eq:basecheck} follows from the
braid-and-measurement instrument; it is not an assumed quartic
measurement.

\subsection{Selecting a different stabilizer parity}

Gaussian permutations of Majoranas, together with known bilinear
corrections, change the syndrome frame. Any nontrivial binary product of
$S_1,S_2,S_3$ can therefore be mapped to the stabilizer tested by the base
circuit. Appendix~\ref{app:affine} gives the sign-resolved construction
and shows that the corresponding discrete frame changes generate the full
affine action on the eight syndrome labels.

We label the tested parity by a nonzero three-bit vector $\ell$. The check
accepts when both input blocks have the same value of this parity. After a
free output-frame correction the syndrome rule is
\begin{equation}
 \begin{split}
 &\text{accept if }\ip{\ell}{r}=\ip{\ell}{s},\\
 &C_{\ell,e}(r,s)=r+s+\ip{\ell}{s}e,
 \qquad \ip{\ell}{e}=1 .
 \end{split}
 \label{eq:check}
\end{equation}
The vector $e$ specifies the output frame. The twenty-eight allowed
pairs $(\ell,e)$ are versions of the same comparison, obtained by
Gaussian frame changes, rather than different physical interactions.
When writing a sequence, we encode the binary labels $001,\ldots,111$
by the integers $1,\ldots,7$. In particular, $1,2,4$ select $S_3,S_2,S_1$,
respectively, and $7$ selects $S_1S_2S_3$. We use the output-frame choices
listed in Appendix~\ref{app:fourier}.

\begin{figure}[t]
\centering
\begin{tikzpicture}[>=Latex,font=\small,
 box/.style={draw,rounded corners=1pt,align=center,inner sep=4pt}]
 \node (r) at (0,0.65) {block 1};
 \node (s) at (0,-0.65) {block 2};
 \node[box] (m) at (1.35,0) {Gaussian\\frame change};
 \node[box] (c) at (3.55,0) {Bravyi\\comparison};
 \node[box] (o) at (5.75,0) {Gaussian\\correction};
 \node (out) at (7.2,0) {output};
 \node (rej) at (3.55,-1.25) {discard};
 \draw[->] (r)--(m.west);
 \draw[->] (s)--(m.west);
 \draw[->] (m)--(c);
 \draw[->] (c)--node[above,yshift=9pt,font=\scriptsize]{agree}(o);
 \draw[->] (o)--(out);
 \draw[->,dashed] (c)--node[right,font=\scriptsize]{disagree}(rej);
\end{tikzpicture}
\caption{One generalized purification check. Gaussian frame changes select
which stabilizer product is compared, while the physical
braid-and-fusion comparison is unchanged. A disagreement is rejected; an
agreement is corrected by Gaussian feed-forward and one block is retained.}
\label{fig:check}
\end{figure}

After a complete round we may also randomize the syndrome frame. Averaging
over the linear frame changes leaves the target population unchanged and
makes the seven nonzero syndrome probabilities equal. We call this
isotropization. It maps any syndrome-diagonal state with target
population $F$ to $\pi_F$ without using a non-Gaussian resource. By mixing
also with a random syndrome translation, the fidelity can be reduced to
any lower value down to the uniform even-sector state. We use this later
to extend an endpoint result to every larger input fidelity.

The calculation can be carried out using the eight syndrome
probabilities or the equivalent seven stabilizer correlators. The
correlators are Fourier coordinates of the probability distribution,
not an additional operation on the quantum state; their update is given
in Appendix~\ref{app:fourier}.
We retain the unequal syndrome populations between comparisons and
isotropize only after a complete round. Physically, this means omitting
the randomization between layers. No knowledge of the individual
block's three stabilizer signs is required.

\section{A four-layer purification round}\label{sec:recurrence}

We first arrange the comparisons in a balanced tree. At each layer,
two independently prepared blocks with the same distribution are
compared. The first layer uses raw ancillas; subsequent layers use the
successful outputs of the preceding layer. A three-layer tree has eight
raw inputs and seven comparison nodes. A four-layer tree has sixteen
raw inputs and fifteen nodes. The syndrome distribution is retained
between layers.

Bravyi's round compares $S_3$, $S_2$, and $S_1$, in that order; we denote
it by $\Rthree$. For the modified round $\Rfour$, two independent
outputs of these three layers are compared once more, before
isotropization. The last layer tests $S_1S_2S_3$. Thus the extra layer
does not act again on the same single block: it compares two processed
blocks, each prepared from eight raw inputs.
Although the product stabilizer supplies no independent syndrome bit,
its comparison rejects wrong outputs which pass the preceding layers.

After a complete round we isotropize and may repeat the same procedure.
For the three-layer recurrence the nontrivial fixed point of the noise
model \eqref{eq:modelb} is
\begin{equation}
 F_{\rm B}=0.61560748080966942\ldots .
 \label{eq:bravyithreshold}
\end{equation}
For the four-layer recurrence we obtain
\begin{equation}
 F_*=0.6047116657215318756305\ldots .
 \label{eq:fstar}
\end{equation}
The map also has the target fixed point at $F=1$ and the uniform fixed
point at $F=1/8$. Its only fixed point between them is $F_*$.
For $F>F_*$, each round increases the fidelity and repeated rounds
converge to the target. We refer to this interval as the purification
basin. Appendix~\ref{app:polynomials} gives the exact recurrence and
the root count establishing this statement. The extra layer therefore
allows a noisier input to be purified, in addition to improving error
suppression close to the target.


\begin{figure}[t]
\centering
\begin{tikzpicture}
\begin{axis}[width=0.98\columnwidth,height=5.4cm,
 xlabel={$F$},ylabel={$R_j(F)-F$},
 xmin=0.5,xmax=1,ymin=-0.25,ymax=0.24,
 tick label style={font=\footnotesize},label style={font=\small},
 legend style={font=\small,draw=none,at={(0.97,0.97)},anchor=north east},
 axis lines=left,xtick={0.5,0.6,0.7,0.8,0.9,1},
 ytick={-0.2,-0.1,0,0.1,0.2}]
 \addplot[black,thin,forget plot] coordinates {(0.5,0)(1,0)};
 \addplot[black,dashed,thick] coordinates {(0.50000,-0.1573246547) (0.50500,-0.1519696366) (0.51000,-0.1463900429) (0.51500,-0.1405962696) (0.52000,-0.1345994914) (0.52500,-0.1284116220) (0.53000,-0.1220452689) (0.53500,-0.1155136822) (0.54000,-0.1088306995) (0.54500,-0.1020106859) (0.55000,-0.0950684704) (0.55500,-0.0880192788) (0.56000,-0.0808786642) (0.56500,-0.0736624353) (0.57000,-0.0663865836) (0.57500,-0.0590672093) (0.58000,-0.0517204484) (0.58500,-0.0443623991) (0.59000,-0.0370090507) (0.59500,-0.0296762137) (0.60000,-0.0223794529) (0.60500,-0.0151340232) (0.61000,-0.0079548095) (0.61500,-0.0008562701) (0.62000,0.0061476150) (0.62500,0.0130433920) (0.63000,0.0198181750) (0.63500,0.0264596838) (0.64000,0.0329562760) (0.64500,0.0392969737) (0.65000,0.0454714853) (0.65500,0.0514702222) (0.66000,0.0572843101) (0.66500,0.0629055963) (0.67000,0.0683266514) (0.67500,0.0735407673) (0.68000,0.0785419512) (0.68500,0.0833249152) (0.69000,0.0878850634) (0.69500,0.0922184751) (0.70000,0.0963218857) (0.70500,0.1001926651) (0.71000,0.1038287938) (0.71500,0.1072288375) (0.72000,0.1103919202) (0.72500,0.1133176963) (0.73000,0.1160063216) (0.73500,0.1184584240) (0.74000,0.1206750735) (0.74500,0.1226577529) (0.75000,0.1244083274) (0.75500,0.1259290160) (0.76000,0.1272223621) (0.76500,0.1282912055) (0.77000,0.1291386547) (0.77500,0.1297680607) (0.78000,0.1301829909) (0.78500,0.1303872046) (0.79000,0.1303846298) (0.79500,0.1301793404) (0.80000,0.1297755350) (0.80500,0.1291775171) (0.81000,0.1283896759) (0.81500,0.1274164689) (0.82000,0.1262624052) (0.82500,0.1249320300) (0.83000,0.1234299107) (0.83500,0.1217606233) (0.84000,0.1199287406) (0.84500,0.1179388210) (0.85000,0.1157953983) (0.85500,0.1135029726) (0.86000,0.1110660022) (0.86500,0.1084888956) (0.87000,0.1057760054) (0.87500,0.1029316221) (0.88000,0.0999599690) (0.88500,0.0968651976) (0.89000,0.0936513837) (0.89500,0.0903225240) (0.90000,0.0868825335) (0.90500,0.0833352428) (0.91000,0.0796843965) (0.91500,0.0759336517) (0.92000,0.0720865768) (0.92500,0.0681466511) (0.93000,0.0641172638) (0.93500,0.0600017146) (0.94000,0.0558032133) (0.94500,0.0515248804) (0.95000,0.0471697478) (0.95500,0.0427407594) (0.96000,0.0382407720) (0.96500,0.0336725566) (0.97000,0.0290387994) (0.97500,0.0243421035) (0.98000,0.0195849899) (0.98500,0.0147698993) (0.99000,0.0098991935) (0.99500,0.0049751573) (1.00000,0.0000000000)};
 \addlegendentry{$\mathcal R_3$}
 \addplot[black,solid,thick] coordinates {(0.50000,-0.2358916878) (0.50500,-0.2291494897) (0.51000,-0.2217863592) (0.51500,-0.2138088308) (0.52000,-0.2052280386) (0.52500,-0.1960598744) (0.53000,-0.1863250786) (0.53500,-0.1760492562) (0.54000,-0.1652628120) (0.54500,-0.1540008025) (0.55000,-0.1423027010) (0.55500,-0.1302120796) (0.56000,-0.1177762101) (0.56500,-0.1050455913) (0.57000,-0.0920734119) (0.57500,-0.0789149598) (0.58000,-0.0656269919) (0.58500,-0.0522670785) (0.59000,-0.0388929381) (0.59500,-0.0255617782) (0.60000,-0.0123296571) (0.60500,0.0007491184) (0.61000,0.0136225480) (0.61500,0.0262414333) (0.62000,0.0385598070) (0.62500,0.0505352882) (0.63000,0.0621293620) (0.63500,0.0733075810) (0.64000,0.0840396918) (0.64500,0.0942996876) (0.65000,0.1040657940) (0.65500,0.1133203934) (0.66000,0.1220498955) (0.66500,0.1302445618) (0.67000,0.1378982926) (0.67500,0.1450083847) (0.68000,0.1515752670) (0.68500,0.1576022224) (0.69000,0.1630951022) (0.69500,0.1680620387) (0.70000,0.1725131620) (0.70500,0.1764603242) (0.71000,0.1799168355) (0.71500,0.1828972149) (0.72000,0.1854169561) (0.72500,0.1874923117) (0.73000,0.1891400962) (0.73500,0.1903775062) (0.74000,0.1912219611) (0.74500,0.1916909601) (0.75000,0.1918019578) (0.75500,0.1915722557) (0.76000,0.1910189096) (0.76500,0.1901586510) (0.77000,0.1890078217) (0.77500,0.1875823207) (0.78000,0.1858975616) (0.78500,0.1839684398) (0.79000,0.1818093082) (0.79500,0.1794339601) (0.80000,0.1768556199) (0.80500,0.1740869376) (0.81000,0.1711399900) (0.81500,0.1680262846) (0.82000,0.1647567678) (0.82500,0.1613418359) (0.83000,0.1577913478) (0.83500,0.1541146401) (0.84000,0.1503205437) (0.84500,0.1464174010) (0.85000,0.1424130844) (0.85500,0.1383150147) (0.86000,0.1341301805) (0.86500,0.1298651571) (0.87000,0.1255261255) (0.87500,0.1211188910) (0.88000,0.1166489021) (0.88500,0.1121212683) (0.89000,0.1075407774) (0.89500,0.1029119131) (0.90000,0.0982388709) (0.90500,0.0935255742) (0.91000,0.0887756895) (0.91500,0.0839926407) (0.92000,0.0791796235) (0.92500,0.0743396180) (0.93000,0.0694754020) (0.93500,0.0645895628) (0.94000,0.0596845083) (0.94500,0.0547624784) (0.95000,0.0498255545) (0.95500,0.0448756701) (0.96000,0.0399146193) (0.96500,0.0349440655) (0.97000,0.0299655501) (0.97500,0.0249804994) (0.98000,0.0199902327) (0.98500,0.0149959685) (0.99000,0.0099988312) (0.99500,0.0049998570) (1.00000,0.0000000000)};
 \addlegendentry{$\mathcal R_4$}
\end{axis}
\end{tikzpicture}
\caption{Fidelity gain of the three- and four-layer recurrences,
with isotropization only after each complete round. The fourth comparison
moves the nontrivial fixed point downward while preserving convergence to
the target above threshold. The curves are shown numerically; the root
count and threshold enclosure are obtained with exact arithmetic.}
\label{fig:gain}
\end{figure}

\subsection{Why the additional comparison gives cubic suppression}
\label{sec:cubic}

Close to the target state the two recurrences are
\begin{align}
 1-R_4(1-\eps)
 &=\frac{384}{343}\eps^3+O(\eps^4),\label{eq:cubic}\\
 1-R_3(1-\eps)
 &=\frac{48}{49}\eps^2+O(\eps^3).
 \label{eq:quadratic}
\end{align}
The exact expansion is given in Appendix~\ref{app:polynomials}. The change
from the second to the third power can also be seen directly from the
syndrome faults.

We call a raw block faulty when its syndrome is nonzero. For a fixed
balanced tree the syndrome rule determines whether the comparisons accept
and which syndrome is left at the output. Table~\ref{tab:leaffaults}
contains the exact low-weight counts. In the four-layer round every
single-fault pattern is rejected. Some two-fault patterns are accepted,
but all of them give the target syndrome. A wrong accepted output appears
for the first time at weight three.

\begin{table}[!tbp]
\caption{Exact low-weight preparation-fault counts in a preassigned tree.
``Correct'' means final syndrome zero. These are faults of raw blocks,
not microscopic faults of the purification hardware.}
\label{tab:leaffaults}
\begin{ruledtabular}
\begin{tabular}{cccc}
Round & Faulty blocks & Correct & Wrong\\
\hline
$\Rthree$ &0&1&0\\
(8 inputs)&1&0&0\\
&2&20&48\\
$\Rfour$&0&1&0\\
(16 inputs)&1&0&0\\
&2&40&0\\
&3&32&384
\end{tabular}
\end{ruledtabular}
\end{table}

To see why the last layer removes the quadratic term, place two faults
among the sixteen raw inputs. If one fault lies in each eight-input
half, a preceding comparison rejects each half. If both lie in the same
half, that half either is rejected, has the correct output, or has a
wrong output with odd syndrome parity. In the last case, comparing
$S_1S_2S_3$ with the fault-free output of the other half rejects the pair.
Thus no two-fault pattern gives a wrong accepted output.
Three-fault patterns can produce an accepted wrong output with even
syndrome parity. There are 384 such assignments, giving the coefficient
in Eq.~\eqref{eq:cubic}.

This counting concerns errors in the sixteen input blocks. It does not
mean that two arbitrary faults in braids, fusion measurements, or
feed-forward are corrected. Operation faults are discussed in
Sec.~\ref{sec:noise}.

\section{Preparing an input for the recurrence}\label{sec:certificate}

The recurrence just described increases the fidelity only above $F_*$.
To use a noisier source, we need an initial sequence that prepares an
output above this boundary. We call this output a seed for the
recurrence. During the initial sequence we retain unequal syndrome
populations. Its fidelity need not increase at every stage, because the
next comparison depends on the full distribution rather than on the
target population alone.

The sequences below were selected by exploring finite words of the
available checks. For the balanced benchmark, both inputs at every stage
were independently prepared copies of the preceding output. For the
heterogeneous construction, we considered a restricted family allowing
one stage to use a fresh raw input. Candidate words were evaluated by
propagating all eight syndrome populations, with entry into the basin
$F_{\rm seed}>F_*$ as the principal criterion. After fixing the displayed
words, we verified their endpoints by rational arithmetic and evaluated
the expected copy costs separately. This search was not exhaustive and
does not establish an optimal length, input threshold, or cost.

\subsection{Balanced-tree benchmark}

We begin with the identical-input construction used for the recurrence.
At every layer, the two inputs are independently prepared copies of the
same intermediate state. One explicit sequence is
\begin{equation}
 W_{25}=(6,1,5,1,2,7,4,5,2)\,(1,6,5,2)^4 ,
 \label{eq:w25}
\end{equation}
where each label specifies the stabilizer parity compared at that layer.
At the isotropic input fidelity $F=0.5997$, it gives
\begin{equation}
 F_{25}=0.60550725738718132769784\ldots>F_* .
 \label{eq:w25result}
\end{equation}
The strict inequality follows from the rational interval calculation in
Appendix~\ref{app:certificate}. We may therefore isotropize the output
and apply the four-layer recurrence.

For comparison, the shorter balanced sequence $W_{14}$ in
Appendix~\ref{app:certificate} gives
\begin{equation}
 W_{14}(\pi_{0.6})(0)=0.6091260418296702\ldots>F_* .
 \label{eq:w14result}
\end{equation}
The cost of $W_{25}$ makes clear the limitation of this construction:
at its stated endpoint, it requires about $1.76\times10^{13}$ raw
copies per seed in expectation. Instead of adding more layers, we now
allow the two inputs at one stage to have different preparation histories.

\subsection{Pairing processed blocks with fresh raw blocks}

The elementary comparison in Eq.~\eqref{eq:check} does not require
identical input distributions. Its action is defined on each pair of
syndrome eigenstates, so two independent mixtures with different
preparation histories can be used in the same circuit. The update in
Appendix~\ref{app:fourier} already includes this case.

Let $\Phi_\ell(P_1,P_2)$ denote the accepted output distribution.
For a fixed raw source $\pi_F$, we distinguish two production steps:
\begin{equation}
 \begin{aligned}
 \mathsf S_\ell(P)&=\Phi_\ell(P,P),\\
 \mathsf R_\ell(P)&=\Phi_\ell(P,\pi_F).
 \end{aligned}
 \label{eq:heterosteps}
\end{equation}
The two occurrences of $P$ in the first line mean separately prepared
blocks, not two uses of the same block. In the second line the raw input
is independent and still has the original fidelity $F$.

Starting from $\pi_F$, apply the following steps from left to right:
\begin{equation}
 \begin{split}
 \mathcal H_{12}={}&
 (\mathsf S_1,\mathsf S_2,\mathsf S_4,\mathsf R_7,
  \mathsf S_1,\mathsf S_3,\\
 &\mathsf S_7,\mathsf S_6,\mathsf S_3,\mathsf S_4,
  \mathsf S_1,\mathsf S_6).
 \end{split}
 \label{eq:h12}
\end{equation}
After the first three stages, a processed block is paired with a fresh
raw block. This fourth stage compares $S_1S_2S_3$, namely the parity of
the three syndrome bits, not the total fermion parity $Q$.
The remaining eight stages again pair independently prepared blocks
with identical distributions. No isotropization is performed between
stages.

The fresh block comes from the same noisy source as every other input.
For example, at raw fidelity $0.5997$, the first three stages give a
processed fidelity of about $0.576884$. Pairing this output with a fresh
raw block raises the fidelity to about $0.582748$, still below the raw
value. The useful change is therefore not captured by fidelity alone:
the subsequent comparisons act on the resulting unequal syndrome
populations. This stage also needs only one processed block rather
than two.

Exact rational evaluation at $F=0.5995$ gives
\begin{equation}
 F_H=0.605026344233701236200599895568\ldots>F_* .
 \label{eq:h12result}
\end{equation}
Appendix~\ref{app:certificate} gives the enclosing rational bounds,
the independent probability-space calculation, and the stage data.
A fully accepted realization has $2304$ raw leaves and $2303$
two-copy comparison nodes. The twelve entries in Eq.~\eqref{eq:h12}
are production stages, not twelve individual circuit operations.
In that tree, the fourth-stage raw-input comparison occurs at $256$
nodes.

\begin{figure*}[!tbp]
\centering
\begin{tikzpicture}[>=Latex, font=\footnotesize,
  schemabox/.style={draw,rounded corners=2pt,align=center,
                     inner sep=5pt},
  group/.style={schemabox,minimum width=2.8cm}]
\node[anchor=west] at (-0.4,3.75) {(a) Four-layer recurrence $\Rfour$};
\node[schemabox] (ra) at (0.9,2.70) {$8$ raw\\blocks};
\node[group] (a3) at (4.15,2.70) {$\ell=1$ ($S_3$)\\
  $\ell=2$ ($S_2$)\\$\ell=4$ ($S_1$)};
\node[schemabox] (rb) at (0.9,0.80) {$8$ raw\\blocks};
\node[group] (b3) at (4.15,0.80) {$\ell=1$ ($S_3$)\\
  $\ell=2$ ($S_2$)\\$\ell=4$ ($S_1$)};
\node[schemabox] (c7) at (8.05,1.75)
    {extra check $\ell=7$\\($S_1S_2S_3$)};
\node[schemabox] (oa) at (11.25,1.75) {one output};
\draw[->] (ra)--(a3);
\draw[->] (rb)--(b3);
\draw[->] (a3.east)--(c7.west);
\draw[->] (b3.east)--(c7.west);
\draw[->] (c7)--(oa);
\node[anchor=west] at (-0.4,-1.15)
  {(b) Heterogeneous initial sequence $\mathcal H_{12}$};
\node[schemabox] (r0) at (0.9,-2.25) {raw\\source};
\node[schemabox] (first) at (3.70,-2.25)
  {$\mathsf S_1,\mathsf S_2,\mathsf S_4$\\($8$ raw leaves per block)};
\node[schemabox] (r7) at (7.10,-2.25)
  {$\mathsf R_7$\\(one fresh input)};
\node[schemabox] (later) at (10.65,-2.25)
  {$\mathsf S_1,\mathsf S_3,\mathsf S_7,\mathsf S_6$\\
   $\mathsf S_3,\mathsf S_4,\mathsf S_1,\mathsf S_6$};
\node[schemabox] (seed) at (14.15,-2.25) {seed};
\node[schemabox] (fresh) at (7.10,-3.75) {fresh raw $\pi_F$};
\draw[->] (r0)--(first);
\draw[->] (first)--(r7);
\draw[->] (fresh)--(r7);
\draw[->] (r7)--(later);
\draw[->] (later)--(seed);
\end{tikzpicture}
\caption{Schematic of the two constructions. (a) Two independently
prepared three-layer outputs are compared with the additional parity
check $\ell=7$, producing a sixteen-input four-layer round. Each
three-layer box represents a balanced eight-to-one tree, not just three
individual circuit nodes. (b) The twelve-stage sequence first makes an
eight-input processed block, then compares it with one fresh source
block at $\mathsf R_7$. The remaining eight stages self-pair
independently prepared copies of the current output. Repeated
subtrees and unsuccessful attempts are suppressed in the drawing.
A fully accepted $\mathcal H_{12}$ tree uses $2304$ raw blocks.}
\label{fig:treearchitectures}
\end{figure*}

Isotropizing this seed and repeating $\Rfour$ then gives arbitrarily
high fidelity, with nonzero success probability for every finite target
accuracy. The result also covers every larger raw fidelity: the free
randomization of Sec.~\ref{sec:checks} can first reduce the source
fidelity to $0.5995$. This applies to all raw blocks, including those
introduced at the fourth stage. We therefore obtain the sufficient
preparation condition
\begin{equation}
 F\geq\frac{5995}{10000}.
 \label{eq:heterothreshold}
\end{equation}
No monotonicity of the intermediate trajectory is assumed.

\begin{corollary}[General even-parity inputs]
\label{cor:eveninputs}
Let $\rho$ be any state of four fermionic modes supported in the even
fermion-parity sector, and let
$F=\bra{a_8}\rho\ket{a_8}$. If independent copies of $\rho$ are
available and $F\geq 5995/10000$, then $\rho$ is
postselection-distillable to $\aEight$ using Gaussian operations,
bilinear measurements, feed-forward, and postselection.
\end{corollary}
\begin{proof}
First apply the random Gaussian unitary channel
\begin{equation}
 \mathcal D(\rho)=\frac18\sum_{v\in\Ftwo^3}
 S^{(v)}\rho S^{(v)}
 =\sum_{s\in\Ftwo^3}\Pi_s\rho\Pi_s .
 \label{eq:syndromedephase}
\end{equation}
Here $S^{(v)}=\prod_{j=1}^3 S_j^{v_j}$, and each $S^{(v)}$
is a Gaussian unitary up to an irrelevant phase: every quartic
$S_j$ is a product of two bilinear Gaussian unitaries.
The identity follows by averaging the characters
$(-1)^{\ip{v}{s}}$; it removes coherences between distinct syndromes
and leaves the target population $F$ unchanged.

Appendix~\ref{app:affine} constructs Gaussian frame changes $U_M$
which implement $s\mapsto Ms$ for $M\in\GL(3,2)$, with no syndrome
translation. Averaging $U_M\mathcal D(\rho)U_M^\dagger$ over this group
fixes $000$ and distributes the other seven populations uniformly,
because $\GL(3,2)$ acts transitively on the nonzero syndromes.
The output is therefore exactly $\sigma_F$ of Eq.~\eqref{eq:modelb}.
For $F>5995/10000$, the Gaussian randomization described in
Sec.~\ref{sec:checks} can reduce its fidelity to $5995/10000$.
Applying the certified initial sequence and then the four-layer
recurrence completes the proof. All randomizations are applied
independently to the input copies.
\end{proof}

On the conditioned depolarized line, Eq.~\eqref{eq:depolarizing}
converts this endpoint to
\begin{equation}
 p\leq\frac{1602}{2551}=0.627989023912191\ldots .
 \label{eq:pbound}
\end{equation}
This remains a result for the even-parity input family, not a strict-FLO
threshold for the unconditioned sixteen-dimensional state.

\section{Resource cost}\label{sec:cost}

We compare expected raw-copy costs, assuming independent preparations and
perfect storage. A failed comparison consumes its two input blocks.
Replacement inputs are prepared independently, and the comparison is
retried. This convention counts rejected attempts but not elapsed time,
memory, or the number of elementary braids in one comparison.

For a balanced depth-$k$ sequence, let $A_j$ be the conditional acceptance
probability at layer $j$. The expected cost obeys $C_0=1$ and
$C_j=2C_{j-1}/A_j$. Its inverse is
\begin{equation}
 Y(W)=2^{-k}\prod_{j=1}^k A_j,\qquad
 4^{-k}\leq Y(W)\leq2^{-k}.
 \label{eq:yield}
\end{equation}
This is also the large-pool yield; a small fixed batch has additional
integer effects from unmatched states.

The probability that all nodes of a single preassigned tree accept is
instead
\begin{equation}
 Z_W=\prod_{j=1}^k A_j^{\,2^{k-j}}.
 \label{eq:treeprob}
\end{equation}
The normalized output state is the same under either convention, but
$Z_W$ is not the output-to-input yield. For $W_{25}$ at $F=0.5997$,
\begin{align}
 \prod_{j=1}^{25}A_j&=1.907457275873\ldots\times10^{-6},\nonumber\\
 Y(W_{25})&=5.68466566763\ldots\times10^{-14}.
 \label{eq:w25yield}
\end{align}

For the heterogeneous schedule the raw-input stage has a different cost:
\begin{equation}
 C_j=
 \begin{cases}
 2C_{j-1}/A_j,&\mathsf S\text{ stage},\\[2pt]
 (C_{j-1}+1)/A_j,&\mathsf R\text{ stage}.
 \end{cases}
 \label{eq:heterocost}
\end{equation}
The additional one counts the fresh raw block. The acceptance
probability is evaluated for these two different inputs.
The retry accounting is analogous to the auxiliary-pair construction
in Eq.~(14) of Ref.~\cite{dur1999}.
Appendix~\ref{app:certificate} derives the formula for the present
comparison and gives an independent expression for the total cost.

\begin{table}[!tbp]
\caption{Costs of the initial sequences with ideal operations. $F$ is the raw fidelity and
$F_{\rm seed}$ the output fidelity before the repeated $\Rfour$ stage.
Values are rounded. The first two rows use the same raw input.
$Y=1/C$ is the inverse expected raw-copy cost.}
\label{tab:bootstrapcost}
\begin{ruledtabular}
\begin{tabular}{ccccc}
Prefix & $F$ & $F_{\rm seed}$ & $C$ & $Y$\\
\hline
$W_{25}$&$0.5997$&$0.605507$&$1.76\,10^{13}$&$5.68\,10^{-14}$\\
$\mathcal H_{12}$&$0.5997$&$0.611888$&$1.04\,10^6$&$9.65\,10^{-7}$\\
$\mathcal H_{12}$&$0.5995$&$0.605026$&$1.05\,10^6$&$9.51\,10^{-7}$
\end{tabular}
\end{ruledtabular}
\end{table}

At the certified heterogeneous endpoint,
\begin{align}
 C(\mathcal H_{12},0.5995)&=1051931.361812\ldots,\nonumber\\
 Y(\mathcal H_{12},0.5995)&=9.506323666185\ldots\times10^{-7}.
 \label{eq:h12cost}
\end{align}
At the common fidelity $0.5997$, the ratio of the heterogeneous yield to
the balanced yield is about $1.69684\times10^7$. The heterogeneous seed
also has a higher fidelity; under ideal free randomization it can be
degraded to the balanced seed fidelity without additional raw-copy cost.
The gain therefore persists when the two constructions are asked to
supply the same seed.

The heterogeneous sequence is much less costly than $W_{25}$, but
roughly one million raw copies per seed remains a large requirement.
Neither sequence has been proved optimal. They also differ in the
later choices of stabilizer products, so the entire gain cannot be
attributed to the fresh-input stage alone.

\subsection{A control for the fresh-input stage}
To isolate the role of the fourth-stage raw block, consider the word
$\mathcal H_{12}^{\rm bal}$ obtained from $\mathcal H_{12}$ by replacing
only $\mathsf R_7$ with $\mathsf S_7$. All twelve comparison labels,
output-frame choices, and subsequent stages are unchanged. The two
constructions therefore differ only in whether the fourth comparison
uses a fresh raw block or a second independently prepared processed
block. Their outputs are compared in Table~\ref{tab:onefreshcontrol}.

\begin{table}[!tbp]
\caption{Controlled comparison of a fresh raw input at stage four
($\mathcal H_{12}$) with self-pairing at that stage
($\mathcal H_{12}^{\rm bal}$). Both columns give the final target
population before isotropization. The recurrence basin begins at
$F_*=0.6047116657\ldots$.}
\label{tab:onefreshcontrol}
\begin{ruledtabular}
\begin{tabular}{ccc}
Raw $F$ & $F_{\rm out}(\mathcal H_{12})$
        & $F_{\rm out}(\mathcal H_{12}^{\rm bal})$\\
\hline
$0.5995$ & $0.6050263442$ & $0.5513873706$\\
$0.5997$ & $0.6118883961$ & $0.5622694793$
\end{tabular}
\end{ruledtabular}
\end{table}

At both source fidelities the heterogeneous word enters the
four-layer purification basin, whereas its all-self-pairing control
does not. The fourth-stage change also lowers the expected raw-copy
cost of this specific twelve-stage sequence; at $F=0.5995$ the costs
are approximately $1.05\times10^6$ and $2.33\times10^6$,
respectively. This control does not establish an advantage over every
possible balanced word. In particular, the larger gain relative to
$W_{25}$ also depends on the remaining check order and word length.

The situation changes after a fixed seed above $F_*$ has been produced.
Because the error is suppressed cubically,
$n=\log_3\log(1/\delta)+O(1)$ rounds are enough to reach infidelity
$\delta$. Each round uses sixteen inputs before postselection. Close to the
target, the product of acceptance probabilities is $1-O(\eps)$, while the
successive infidelities decrease cubically. The total acceptance penalty
over the later rounds is therefore bounded. We obtain
\begin{equation}
 N_{\rm seed}(\delta)
 =O\!\left([\log(1/\delta)]^{\log_3 16}\right).
 \label{eq:overhead}
\end{equation}
The exponent is $\log_3 16=2.5237\ldots$, instead of $\log_2 8=3$ for
$\Rthree$ \cite{bravyi2006}. A derivation of the bounded acceptance
penalty is included in Appendix~\ref{app:factorycomparison}.

This is an expected input-copy cost at fixed seed fidelity, rather than
a space-time cost for an error-corrected device. The distinction also
arises for magic-state factories
\cite{campbell2017,ogorman2017,litinski2019}.
The initial preparation multiplies this high-accuracy cost by a constant
for a fixed seed. That constant can be large near the sufficient
preparation endpoint, even for the heterogeneous sequence.

\subsection{Comparison at a fixed output accuracy}

The initial sequence prepares a seed; the recurrent rounds improve its
accuracy. Table~\ref{tab:architecture} gives the resources of one fully
accepted construction of each type. It excludes rejected attempts,
raw-state preparation, parity selection, and Gaussian frame compilation.
The depth counts comparison layers, not elementary gates or elapsed
time in a device with retries and finite memory.

\begin{table}[!tbp]
\caption{Resources of one fully accepted construction. The last column
is an initial sequence; the first two are recurrent purification rounds.
Each comparison uses four bilinear fusion measurements.}
\label{tab:architecture}
\begin{ruledtabular}
\begin{tabular}{lrrr}
 & $\Rthree$ & $\Rfour$ & $\mathcal H_{12}$\\
\hline
Input leaves & 8&16&2304\\
Comparison nodes &7&15&2303\\
Bilinear measurements &28&60&9212\\
Comparison depth &3&4&12
\end{tabular}
\end{ruledtabular}
\end{table}

To compare complete protocols, we fix both the raw fidelity and the
required output accuracy. We grant elementary-comparison pooling and
independent retry to all constructions, including $\Rthree$.
This differs from using the all-nodes-accept probability in Bravyi's
original whole-round resource expression \cite{bravyi2006}.
For each prescribed construction, we stop at the first complete round
with output infidelity at most $10^{-6}$.
These are numerical examples, not optimized costs.
Table~\ref{tab:matchedcost} includes the final errors because different
whole-round schedules overshoot the same target by different amounts.

\begin{table*}[!tp]
\caption{Numerical expected raw-copy costs at target infidelity
$10^{-6}$, with ideal operations and storage. Each construction uses
the same elementary-comparison retry convention. $n$ is the number of
recurrent rounds, excluding the heterogeneous initial sequence when present.
The full prefix is used when specified; at $F=0.9$ it already meets
the target. No optimization over mixed rounds or partial prefixes
has been performed.}
\label{tab:matchedcost}
\begin{ruledtabular}
\begin{tabular}{clrrr}
Raw $F$&Construction&$n$&Expected raw copies&Final infidelity\\
\hline
$0.65$&$\Rthree$ only&6&$1.01149\,10^7$&$4.56662\,10^{-9}$\\
$0.65$&$\Rfour$ only&4&$1.50305\,10^6$&$1.32397\,10^{-11}$\\
$0.65$&$\mathcal H_{12}$ then $\Rfour$&1&$1.11976\,10^6$&$2.56970\,10^{-9}$\\
$0.9$&$\Rthree$ only&3&$769.941$&$3.00210\,10^{-8}$\\
$0.9$&$\Rfour$ only&2&$379.157$&$6.16158\,10^{-9}$\\
$0.9$&full $\mathcal H_{12}$&0&$3918.263$&$2.88669\,10^{-15}$
\end{tabular}
\end{ruledtabular}
\end{table*}

At $F=0.65$ the heterogeneous construction followed by $\Rfour$
uses fewer raw copies than either repeated recurrence alone in this
comparison. At $F=0.9$ the full initial sequence is unnecessary and its cost
is higher. The choice of prefix therefore depends on the source
fidelity and required final accuracy.

Near the sufficient endpoint, most of the cost can occur after the seed
has been prepared. At $F=0.5995$, the stated heterogeneous sequence
followed by eight complete $\Rfour$ rounds requires about
$2.78\times10^{20}$ raw copies in expectation to meet the same target.
The seed lies only slightly above the unstable fixed point, where the
recurrence improves it slowly. This cost belongs to the prescribed
schedule and is not a lower bound for all purification protocols.
Appendix~\ref{app:factorycomparison} specifies the calculation and
stopping rule.

\subsection{Physical implementation and use of the output}

At the level of one comparison, the proposed modification requires no
new interaction. Two eight-Majorana blocks undergo Gaussian frame
changes and Bravyi's inter-register braid network. Four bilinear fusion
outcomes then determine acceptance and the Gaussian correction
\cite{bravyi2006}. The fourth recurrence layer adds another comparison
of processed blocks; the heterogeneous stage instead changes where one
of the two input blocks comes from.

For the computation, the purified state supplies an entangling Clifford
gate in Bravyi's encoding. A separate four-Majorana phase ancilla is
needed for the non-Clifford operation \cite{bravyi2006}.
Reducing the number of raw eight-Majorana ancillas therefore reduces the
preparation demand for this part of the scheme. It does not by itself
reduce every contribution to the cost of universal computation. The
state is also a resource for FLO computations and provides a concrete
mixed-state conversion problem for fermionic resource theories.

Proposals using branched semiconductor--superconductor structures,
charging-energy control, and parity readout provide possible routes to
the required primitives \cite{aasen2016}.
Majorana-island architectures offer related control and readout methods
\cite{karzig2017}. Applying the present schedules in either setting
would still require a layout and a compilation of the comparisons.
The expected raw-copy count does not determine peak storage, waiting-time
errors, parity leakage, or the number of exchanges. Moreover, an
architecture that grants joint multi-Majorana measurements has a
different free-operation set from the bilinear measurements assumed
here. These implementation questions remain separate from the
purification bounds.

\section{Robustness to imperfect purification operations}
\label{sec:noise}

So far, only the input ancillas have been noisy. We now ask what happens
when the comparisons themselves are imperfect. First, we add an
effective syndrome error after each accepted comparison. We then
consider readout errors and faulty accepted branches. These models do
not resolve individual hardware processes; they show which conclusions
depend on ideal purification operations.

\subsection{Effective syndrome noise and the error floor}

Suppose that after a successful comparison the retained block acquires an
independent random syndrome shift $d$ with distribution
\begin{equation}
 \nu_\lambda(d)=
 \begin{cases}
 1-\lambda,&d=0,\\
 \lambda/7,&d\ne0 .
 \end{cases}
 \label{eq:noise}
\end{equation}
The state remains in the even syndrome-diagonal sector. Applying this
channel after every successful comparison defines the noisy recurrence
$R_{4,\lambda}$. Its exact form in stabilizer-correlator coordinates is
given in Appendix~\ref{app:noise}.

Before solving the full recurrence, one can already see the first-order
effect from a single fault. A successful $\Rfour$ tree contains fifteen
comparison nodes. Table~\ref{tab:operationfaults} gives the result of
inserting one nonzero syndrome fault after one node, starting from sixteen
perfect input blocks. Faults in the first layer are always found by later
comparisons. The closer a fault occurs to the output, the fewer checks
remain, and a fault after the last comparison cannot be detected.

\begin{table}[!tbp]
\caption{One nonzero syndrome fault inserted after an otherwise ideal
check of the $\Rfour$ tree, with sixteen perfect input blocks.
All locations and nonzero labels are distinguished.}
\label{tab:operationfaults}
\begin{ruledtabular}
\begin{tabular}{crrr}
Layer & Locations & Accepted, wrong & Rejected later\\
\hline
1&8&0&56\\
2&4&4&24\\
3&2&6&8\\
4&1&7&0\\
Total&15&17&88
\end{tabular}
\end{ruledtabular}
\end{table}

There are 105 labeled one-fault events. Eighty-eight are rejected and
seventeen are accepted with a wrong syndrome. Hence, for perfect inputs,
\begin{align}
 Z_{4,\lambda}(1)&=1-\frac{88}{7}\lambda+O(\lambda^2),\nonumber\\
 1-R_{4,\lambda}(1)&=\frac{17}{7}\lambda+O(\lambda^2).
 \label{eq:opfirstorder}
\end{align}
Thus the cubic suppression of preparation errors does not imply cubic
suppression of operation faults. The latter enter linearly. Since the ideal
map has zero slope at the target fixed point, the attracting fixed point is
shifted to
\begin{equation}
 F_+(\lambda)=1-\frac{17}{7}\lambda+O(\lambda^2).
 \label{eq:floor}
\end{equation}
Repeated purification therefore approaches a state with nonzero error
when $\lambda$ is fixed. For $\Rthree$, the same calculation gives the
coefficient $17/7$. Although the fourth layer adds comparison nodes,
later checks reject enough earlier faults to leave the leading
output-error coefficient unchanged. This equality does not extend to
the acceptance probability or the size of the purification basin.
Appendix~\ref{app:noise} gives the fault counts and the fixed-point
argument. It also shows directly why noise after the last comparison,
with no further error information, sets a linear error floor.

\subsection{Purification boundaries and sensitivity of the initial sequence}

The lower fixed point also moves. Expanding around the ideal unstable
fixed point gives
\begin{equation}
 F_-(\lambda)=F_*+1.736536\ldots\,\lambda+O(\lambda^2).
 \label{eq:lowerbranch}
\end{equation}
Therefore the purification interval shrinks from both sides. Numerical
continuation of this effective model shows that the two fixed points meet
at approximately $\lambda=4.75\times10^{-2}$ and $F=0.787$. The numerical
data and tangency calculation are collected in Appendix~\ref{app:noise}.
These values characterize only the effective syndrome-noise model and are
not device thresholds.

The long initial sequence is much more sensitive. In the ideal case,
$W_{25}$ applied to $F=0.5997$ ends only about $8.0\times10^{-4}$ above
the basin boundary. If the channel \eqref{eq:noise} is inserted after
every comparison, this margin disappears at
\[
 \lambda\simeq1.0\times10^{-7}.
\]
For $W_{14}$ applied at $F=0.6$, the corresponding value is
\[
 \lambda\simeq2.5\times10^{-5}.
\]
The shorter sequence therefore tolerates an effective error about 250 times
larger, while giving up only $3\times10^{-4}$ in the ideal input fidelity.
The two rows use different input fidelities, so this is not a
same-source comparison.

At the common raw fidelity $F=0.5997$, the heterogeneous initial sequence has a
local numerical crossing at approximately $1.29318\times10^{-4}$,
about 1284 times the $W_{25}$ value. At its stronger certified endpoint
$F=0.5995$, the crossing is approximately $5.65512\times10^{-6}$.
These are local numerical crossings in the effective syndrome-noise
model, not certified global thresholds. At the same raw fidelity, the
heterogeneous sequence has a much larger tolerance in this model;
moving its input closer to the sufficient endpoint greatly reduces
that tolerance.

\subsection{Readout faults and more general faulty branches}

The syndrome channel above does not describe an incorrect classical
measurement record. Let the true fusion outcomes in one elementary
comparison be $t_j$, while the recorded outcomes are $t_j+m_j$. On a
genuinely accepted branch, the error in the feed-forward frame is
\begin{equation}
 \delta_{\rm ff}=(m_1+m_2,\ m_2+m_3,\ 0).
 \label{eq:readouterror}
\end{equation}
On a truly accepted branch, one readout-bit flip changes the checksum
and leads to rejection. Two flips keep the checksum unchanged but produce
a nonzero residual frame shift; four flips cancel in the output frame.
Let $\eps_{\rm ro}$ denote the probability of a nonzero readout-induced
shift, conditioned on both true and recorded acceptance. If the four
readout bits flip independently with probability $\mu$, then
\begin{equation}
 \eps_{\rm ro}(\mu)=
 \frac{6\mu^2(1-\mu)^2}
 {(1-\mu)^4+6\mu^2(1-\mu)^2+\mu^4}
 =6\mu^2+O(\mu^3).
 \label{eq:readout}
\end{equation}
For otherwise perfect inputs this is also the output infidelity.
For an already erroneous output, a residual shift can cancel an existing
syndrome, so Eq.~\eqref{eq:readout} is not its total infidelity.
False acceptance of a pair which should have been rejected is a different
effect and depends on the full noisy measurement instrument.

A simple worst-case estimate is possible if the ideal branch keeps its
ideal statistics. Suppose that the accepted instrument has the form
\begin{equation}
 \mathcal E_{\rm acc}
 =(1-t)\mathcal E_{\rm acc}^{\rm id}
   +t\mathcal E_{\rm acc}^{\rm bad},\qquad
 0\leq t\leq\eta_{\rm tree},
 \label{eq:mixture}
\end{equation}
where the bad map is completely positive and trace nonincreasing. No
restriction is imposed on the output of the bad branch. Then
\begin{equation}
 F_{\rm out}\geq
 \frac{(1-\eta_{\rm tree})Z_4(F)R_4(F)}
 {(1-\eta_{\rm tree})Z_4(F)+\eta_{\rm tree}}.
 \label{eq:badbound}
\end{equation}
Appendix~\ref{app:noise} gives the derivation and explains why the
ideal-branch assumption is needed. For fifteen independent fault
locations, each failing with probability $\eta$, one may take
$\eta_{\rm tree}=1-(1-\eta)^{15}$. With correlated fault flags, the union
bound gives $\eta_{\rm tree}\leq15\eta$ if the same branch decomposition
remains valid.

At the ideal $W_{25}$ seed, asking only that the bound
\eqref{eq:badbound} improve the seed gives
$\eta_{\rm tree}\lesssim2.75\times10^{-6}$, or
$\eta\lesssim1.83\times10^{-7}$ for fifteen independent identical
locations. These values are intentionally conservative and should not be
identified with microscopic hardware thresholds.

There is one more limitation. The elementary comparison detects a
difference between the two input syndromes. A common syndrome shift of
both inputs can pass through the check. The exact identity is given in
Appendix~\ref{app:noise}. For this reason the independent input-noise model
cannot be extended to arbitrary correlated errors without further
assumptions.

\section{Discussion and conclusions}\label{sec:discussion}

We have considered how the non-Gaussian ancilla in Bravyi's Majorana
computation scheme can be purified more efficiently using the same
elementary braid-and-fusion comparison. The main change is to compare
the product of the three stabilizers after the first three layers.
Although this product adds no independent syndrome bit, the comparison
detects the wrong two-block preparation patterns that survive those
layers. The first harmful accepted patterns then contain three faulty
raw blocks. This changes the error suppression from quadratic to cubic
and reduces the asymptotic raw-copy overhead.

The second change concerns inputs for which this recurrence does not
yet improve the fidelity. By pairing a processed block with a fresh
copy from the same source, we obtain an initial sequence that prepares
a seed at raw fidelity $F=0.5995$. At the common fidelity $0.5997$, it
requires more than seven orders of magnitude fewer raw copies than the
balanced benchmark and has a larger tolerance to the effective
operation noise. No additional non-Gaussian operation or cleaner source
is used.

These improvements address ancilla preparation rather than the complete
cost of a quantum computation. In Bravyi's scheme the purified state
supports an entangling Clifford gate, while the non-Clifford resource
must be supplied separately. The finite-accuracy comparisons also show
why a lower preparation bound is not enough to assess usefulness: a
full initial sequence may be unnecessary for a good source, whereas a
seed just above the unstable fixed point can remain very expensive to
purify. With imperfect comparisons, the additional layer leaves the
leading operation-error coefficient unchanged in the model considered
here, but the final noise still prevents convergence to a pure state.

It is useful to compare the achieved preparation fidelity with the
boundary of the free states. For the family \eqref{eq:modelb}, the
four-mode concurrence criterion \cite{oszmaniec2014,uhlmann2000} gives
\begin{equation}
 C_+(\sigma_F)=
 \max\left\{0,\,2\max\left(F,\frac{1-F}{7}\right)-1\right\}.
 \label{eq:concurrence}
\end{equation}
Thus $\sigma_F$ is convex-Gaussian exactly when $F\leq1/2$.
FLO preserves the convex-Gaussian set, including on any postselected
branch with nonzero probability, so these inputs cannot be distilled
to $\aEight$. Together with the present construction, this gives
\begin{equation}
 \begin{array}{ll}
 F\leq1/2:&\text{not distillable by FLO},\\
 F\geq0.5995:&\text{postselection-distillable}.
 \end{array}
 \label{eq:regions}
\end{equation}
The interval $1/2<F<0.5995$ remains unresolved. The value $0.5995$ is a
sufficient input fidelity for the stated construction, not an optimal
threshold over all FLO protocols. Corollary~\ref{cor:eveninputs}
extends the sufficient part of this statement beyond isotropic states
to all independent even-parity sources with the same target overlap.

This interval is related to the transition region in noisy Majorana
computation discussed in Ref.~\cite{demelo2013}: states can lie outside
the convex-Gaussian set without a known distillation procedure.
Fermionic convolution, antiflatness, and two-copy measures provide
other ways to characterize non-Gaussianity
\cite{coffman2025,sierant2025,haug2026,tarabunga2026}.
Whether these tools can help determine distillability of the remaining
mixed states is a further question.

Two extensions are particularly natural. More general heterogeneous
trees could be optimized for a specified source fidelity, output
accuracy, and operation-noise level, rather than for the lowest input
fidelity alone. One could also allow continuous Gaussian rotations and
intermediate measurements beyond the discrete frame changes used here.
Both extensions would retain the same supplied non-Gaussian resource.
The present results show that its preparation can already be improved
by changing how the available comparisons are combined.

\section*{Data and code availability}
No experimental data were used. The syndrome update, comparison words,
recurrence coefficients, and rational enclosures are given in the paper.
The accompanying archive (available from the authors upon request) \texttt{majorana\_verification.zip} contains
Python scripts for the integer and rational certificates, the
independent recurrence checks, the controlled fourth-stage comparison,
and the numerical resource and effective-noise calculations. A README
specifies the dependencies and commands needed to reproduce these
calculations. The noise-continuation values are numerical, whereas the
stated noiseless basin-entry bounds are certified by exact arithmetic.

\begin{acknowledgments}
This work was carried out under the IRA Programme, project
No.~FENG.02.01-IP.05-0006/23, financed by the FENG Programme 2021--2027,
Priority FENG.02, Measure FENG.02.01, with support from the FNP. The authors also acknowledge the Wrocław Centre for Networking and Supercomputing (WCSS) for providing computational resources used in numerical and symbolic verification and testing of the accompanying scripts. ChatGPT
(OpenAI) was used for proof-checks of derivations and as a help in creating numerical scripts, language editing and for improving the
presentation and exposition of the manuscript.  The authors are
responsible for the scientific content. 
\end{acknowledgments}

\appendix

\section{Affine labels and the Majorana-permutation normalizer}
\label{app:affine}

For $v\in\Ftwo^3$, write $S^{(v)}=S_1^{v_1}S_2^{v_2}S_3^{v_3}$.
For an ordered support $A=\{a_1<\cdots<a_m\}$, put
$c_A=c_{a_1}\cdots c_{a_m}$. The Majorana algebra gives
\begin{equation}
 \begin{split}
 c_A^\dagger&=(-1)^{m(m-1)/2}c_A,\\
 c_A^2&=(-1)^{m(m-1)/2}I,\\
 c_Ac_B&=(-1)^{|A||B|-|A\cap B|}c_Bc_A .
 \end{split}
 \label{eq:signrules}
\end{equation}
In particular, a single Majorana anticommutes with a quartic monomial
precisely when it belongs to its support. Define the label of $c_p$ by its
three commutation signs with $S_1,S_2,S_3$. In Majorana-index order,
\begin{equation}
 \bigl(\lambda(p)\bigr)_{p=1}^8
 =(101,111,011,001,100,110,010,000).
 \label{eq:labels}
\end{equation}
This table follows directly from Eq.~\eqref{eq:stabilizers}.

For $v\ne0$, the support of $S^{(v)}$ is the four-point set
\[
 A_v=\{p:\ip{v}{\lambda(p)}=1\}.
\]
Products correspond to symmetric differences of supports.
Multiplication by $Q$ complements a support, so the unsigned supports
of $\{S^{(v)},QS^{(v)}:v\ne0\}$ are exactly the fourteen affine planes
of $\Ftwo^3$. Their incidence vectors, together with the empty and
full supports, form the binary first-order Reed--Muller code
$\mathrm{RM}(1,3)$ \cite{macwilliams1977}.

Every nonzero translation $w$ is realized by a bilinear unitary.
Choose distinct $p,q$ with $\lambda(p)+\lambda(q)=w$. Then
\[
 V_w=c_pc_q=\exp\!\left(\frac{\pi}{2}c_pc_q\right)
\]
preserves parity and has commutation character $w$ with the three
stabilizers. Hence $V_w\Pi_sV_w^\dagger=\Pi_{s+w}$.

Now take $M\in\GL(3,2)$ and let $\pi_M$ permute the Majoranas according
to $\lambda(\pi_M(p))=M\lambda(p)$. A swap of two label coordinates
acts on the eight points as two disjoint transpositions. The same
is true of a transvection $x_i\mapsto x_i+x_j$.
These elementary transformations generate $\GL(3,2)$, so every
$\pi_M$ is even. The Gaussian implementation of $\SO(8)$
\cite{bravyi2005} therefore supplies a spin implementer $U_M$ of its
orientation-preserving permutation matrix satisfying
$U_Mc_pU_M^\dagger=c_{\pi_M(p)}$.

A plane normal transforms contragrediently. Including the monomial
reordering sign, this gives, on the even sector,
\begin{equation}
 U_MS^{(v)}U_M^\dagger
 =\eta_M(v)S^{(M^{-T}v)},\qquad \eta_M(v)\in\{1,-1\}.
 \label{eq:covariance}
\end{equation}
Because $S^{(v+w)}=S^{(v)}S^{(w)}$, the signs satisfy
$\eta_M(v+w)=\eta_M(v)\eta_M(w)$. There is a unique $b_M$ with
$\eta_M(v)=(-1)^{\ip{b_M}{v}}$. If the transformed projector has
syndrome $t$, Eq.~\eqref{eq:covariance} implies
$\ip{M^{-T}v}{t}=\ip{v}{s+b_M}$ for every $v$.
Thus $t=M(s+b_M)$. Applying $V_{Mb_M}$ after $U_M$ cancels the
offset and realizes exactly $s\mapsto Ms$.

Combining the translations with the linear frame changes realizes every
affine syndrome transformation
\begin{equation}
 s\longmapsto Ms+w,\qquad M\in\GL(3,2),\quad w\in\Ftwo^3 .
 \label{eq:affine}
\end{equation}
Their number is $8(8-1)(8-2)(8-4)=1344$.
The availability of the linear subgroup is also already implicit in
Bravyi's explicit syndrome shears \cite{bravyi2006}:
\[
 \begin{aligned}
 T_{12}:s&\mapsto(s_1+s_2,s_2,s_3),\\
 T_{23}:s&\mapsto(s_1,s_2+s_3,s_3),\\
 T_{31}:s&\mapsto(s_1,s_2,s_3+s_1).
 \end{aligned}
\]
They generate $\GL(3,2)$. The label construction above supplies a
uniform sign-resolved description of the same freedom.

For completeness, suppose an unsigned Majorana permutation preserves
the set of quartic stabilizers and their complements. The corresponding
bijection $f$ of $\Ftwo^3$ preserves every affine plane.
Put $g(x)=f(x)+f(0)$. For distinct nonzero $x,y$ the plane
$\{0,x,y,x+y\}$ is mapped to a plane containing
$0,g(x),g(y)$. Its fourth point must be $g(x)+g(y)$, so
$g(x+y)=g(x)+g(y)$. The identity also holds if one vector is zero or
the two vectors coincide. Therefore $g$ is invertible and linear,
and $f$ is affine. The converse follows immediately by mapping plane
equations. This proves the stated unsigned normalizer, without
making a claim about continuous Gaussian rotations.

\section{Syndrome coordinates and the exact check update}
\label{app:fourier}

We express the recurrence in terms of the eight syndrome probabilities
or, equivalently, seven stabilizer correlators. For
$v\in\Ftwo^3$ define
\begin{equation}
 \begin{aligned}
 x_v(P)&=\sum_s(-1)^{\ip{v}{s}}P(s),\\
 P(s)&=\frac18\sum_v(-1)^{\ip{v}{s}}x_v(P).
 \end{aligned}
 \label{eq:fourier}
\end{equation}
These are not new observables: if
$S^{(v)}=S_1^{v_1}S_2^{v_2}S_3^{v_3}$, then
$x_v=\Tr[S^{(v)}\rho_P]$. In particular,
\begin{equation}
 \begin{aligned}
 x_0&=1,\qquad P(0)=\frac18\sum_vx_v,\\
 x_v(\pi_F)&=\frac{8F-1}{7}\qquad(v\ne0).
 \end{aligned}
 \label{eq:fidelity}
\end{equation}

For recurrence calculations we encode the seven nonzero binary labels
$001,\ldots,111$ by the integers $1,\ldots,7$. A convenient canonical
choice of output-frame correction is
\begin{equation}
 (e(1),\ldots,e(7))=(1,2,1,4,1,2,1).
 \label{eq:sections}
\end{equation}
Other admissible choices describe the same physical comparison in a
different free output frame.

For independent inputs $P_1,P_2$, let
$x_v^{(i)}=x_v(P_i)$ and
$D=1+x_\ell^{(1)}x_\ell^{(2)}$. The acceptance probability of the
generalized check is
\begin{equation}
 A_{\ell}=\frac{D}{2}.
 \label{eq:acceptance}
\end{equation}
Conditioned on acceptance, the correlators update as
\begin{equation}
 x'_v=
 \begin{cases}
 \dfrac{x_v^{(1)}x_v^{(2)}
       +x_{v+\ell}^{(1)}x_{v+\ell}^{(2)}}{D},
       &\ip{v}{e}=0,\\[6pt]
 \dfrac{x_v^{(1)}x_{v+\ell}^{(2)}
       +x_{v+\ell}^{(1)}x_v^{(2)}}{D},
       &\ip{v}{e}=1 .
 \end{cases}
 \label{eq:update}
\end{equation}
To derive this expression, write
$\chi_v(s)=(-1)^{\ip{v}{s}}$ and
\[
 {\bf1}_{\rm acc}(r,s)
 =\tfrac12[1+\chi_\ell(r)\chi_\ell(s)] .
\]
On an accepted pair,
$\chi_v(C_{\ell,e}(r,s))
=\chi_v(r)\chi_v(s)\chi_\ell(s)^{\ip{v}{e}}$.
Expanding the acceptance indicator and averaging over the independent
input distributions gives Eqs.~\eqref{eq:acceptance} and
\eqref{eq:update}.

For the parity being checked,
\begin{equation}
 x'_\ell=\frac{x_\ell^{(1)}+x_\ell^{(2)}}
                   {1+x_\ell^{(1)}x_\ell^{(2)}} .
 \label{eq:checked}
\end{equation}
Thus agreement reinforces the tested stabilizer parity, but not all
seven correlators simultaneously. This is the reason the ordering of
comparisons matters. For identical inputs the acceptance probability is
at least one half, so every finite identical-input word has strictly
positive success probability.

\paragraph{Realization of the generalized comparison.}
Write $a=001$. Applying a linear syndrome map $M$ on both inputs,
the base comparison of Appendix~\ref{app:circuit}, and $M^{-1}$
on the output gives Eq.~\eqref{eq:check} with
$\ell=M^Ta$ and $e=M^{-1}a$.
Conversely, fix $\ip{\ell}{e}=1$ and an isomorphism
$L:\ker\ell\to\ker a$. The decomposition
$x=\ip{\ell}{x}e+k$, $k\in\ker\ell$, is unique. Define
$Mx=\ip{\ell}{x}a+Lk$.
This map is invertible, satisfies $Me=a$, and obeys
$\ip{a}{Mx}=\ip{\ell}{x}$ for every $x$; hence it has the required
$\ell$ and $e$. The frame realization in Appendix~\ref{app:affine}
therefore implements all seven choices of $\ell\ne0$ and the four
admissible $e$ for each, using the same physical instrument.

\paragraph{Positivity for the heterogeneous schedule.}
Equation~\eqref{eq:update} requires
$1+x_\ell^{(1)}x_\ell^{(2)}>0$. Unlike the identical-input case,
arbitrary unequal inputs need not satisfy this: two deterministic
syndromes with different checked parities have zero acceptance.
For the schedule in Eq.~\eqref{eq:h12}, however, positivity follows
without a numerical assumption. At $F\ge1/8$ every raw nontrivial
correlator is nonnegative. Each numerator in Eq.~\eqref{eq:update}
is a sum of nonnegative products, and its denominator is at least one
whenever both inputs have nonnegative correlators. This property is
preserved at self-pairing and fresh-raw stages. Induction therefore gives
$A_j\ge1/2$ throughout this particular family of schedules.

\section{Bravyi's comparison and its classical controls}
\label{app:circuit}

We recall the braid-and-fusion comparison from Ref.~\cite{bravyi2006}.
The two blocks occupy $c_1,\ldots,c_8$ and
$c_9,\ldots,c_{16}$. Bravyi's inter-register braid network $B$, shown
in Fig.~2 of Ref.~\cite{bravyi2006}, is followed by the four bilinear
measurements
\begin{equation}
 T_j=-ic_{2j+7}c_{2j+8},\qquad j=1,\ldots,4,
 \label{eq:fusions}
\end{equation}
with eigenvalues $(-1)^{t_j}$. Its checksum identity is
\begin{equation}
 B^\dagger(T_1T_2T_3T_4)B
 =(c_5c_6c_7c_8)(c_9c_{10}c_{11}c_{12}).
 \label{eq:checksumidentity}
\end{equation}
On even input blocks the right-hand side has eigenvalue
$(-1)^{r_3+s_3}$. Retaining only
$t_1+t_2+t_3+t_4=0$ therefore enforces $r_3=s_3$.

Use the exchange convention
$B_{p,q}=\exp(-\pi c_pc_q/4)$ and
$D_{p,q}=B_{p,q}^2=-c_pc_q$. The accepted branch is corrected by
\begin{equation}
 D_{2,3}^{\,t_1+t_2+1}D_{1,2}^{\,t_2+t_3+1}.
 \label{eq:feedforward}
\end{equation}
All exponents are binary. The exchange identities and output
stabilizer relations in Eqs.~(25), (27), and (28) of
Ref.~\cite{bravyi2006} yield
$u_1=r_1+s_1$, $u_2=r_2+s_2$, and $u_3=r_3=s_3$.
These output relations, together with the checksum, give
Eq.~\eqref{eq:basecheck}. We use the circuit identity from
Ref.~\cite{bravyi2006}; compilation into a particular nearest-neighbor
layout is not part of the present calculation.

The readout formula can be checked directly in this convention.
The label difference $\lambda(2)+\lambda(3)=100$ shows that
$D_{2,3}$ flips the first syndrome bit, while
$\lambda(1)+\lambda(2)=010$ shows that $D_{1,2}$ flips the second.
Replacing $t_j$ by $t_j+m_j$ changes the two control bits by
$m_1+m_2$ and $m_2+m_3$, respectively, giving
Eq.~\eqref{eq:readouterror}. Of the six weight-two masks, two give
each of $100$, $010$, and $110$.
Together with the checksum parity, this proves
Eq.~\eqref{eq:readout} as the probability of a nonzero residual
shift conditional on true and recorded acceptance. It is the output
infidelity for otherwise perfect inputs. It does not count error
cancellation on imperfect inputs or falsely accepted, initially
mismatched quantum branches.

\section{Exact recurrence polynomials and fault enumerators}
\label{app:polynomials}

A useful reconstruction works entirely in probability space.
For each syndrome $s$, let $H_s^{(j)}(z)$ count accepted assignments
of a depth-$j$ subtree, with the power of $z$ equal to the number of
faulty leaves. Initially,
\begin{equation}
 H_0^{(0)}(z)=1,\qquad H_s^{(0)}(z)=z\quad(s\ne0).
 \label{eq:initpolys}
\end{equation}
At layer $j$ use the exact integer-polynomial recursion
\begin{equation}
 H_u^{(j)}
 =\!\!\sum_{\substack{\ip{\ell_j}{r}=\ip{\ell_j}{s}\\
 r+s+\ip{\ell_j}{s}e(\ell_j)=u}}
 H_r^{(j-1)}H_s^{(j-1)}.
 \label{eq:countrecursion}
\end{equation}
The two factors refer to distinct subtrees, so their product counts
all paired assignments once. The acceptance constraint excludes
any assignment rejected at the new node; rejections below it have
already been excluded by the preceding polynomials.

Write $c_w=[z^w]H_0^{(k)}$ and
$d_w=\sum_{s\ne0}[z^w]H_s^{(k)}$.
For $n=2^k$ leaves and $q=(1-F)/7$,
\begin{align}
 G_W(F)&=\sum_{w=0}^n c_wF^{n-w}q^w,\nonumber\\
 B_W(F)&=\sum_{w=0}^n d_wF^{n-w}q^w,\nonumber\\
 Z_W(F)&=G_W(F)+B_W(F),\qquad
 R_W(F)=\frac{G_W(F)}{Z_W(F)}.
 \label{eq:probpolys}
\end{align}
Here $G_W$ and $B_W$ are the unnormalized accepted correct and wrong
probabilities. This calculation neither uses Fourier coordinates
nor stores a precomputed rational recurrence.

For $\Rfour$, the full coefficients are given in
Table~\ref{tab:coeffs}. For the four-layer recurrence the fixed-point equation factorizes as
\begin{equation}
 R_4(F)-F
 =-\frac{(F-1)(8F-1)g_{15}(F)}{D_4(F)} .
 \label{eq:factorization}
\end{equation}
In this normalization,
\begin{equation}
 D_4(F)=7^{16}Z_{\Rfour}(F).
 \label{eq:denominator}
\end{equation}
It is strictly positive throughout $[0,1]$: the all-correct term is
positive for $F>0$, and the nonzero weight-16 coefficient gives
positivity at $F=0$.
The table also defines $g_{15}(F)=\sum_{j=0}^{15}g_jF^j$.
Substitution into Eq.~\eqref{eq:probpolys} verifies the
factorization by an exact polynomial identity.

\begin{table*}[!tp]
\caption{Complete integer data for the four-layer recurrence.
The first three columns give the correct and wrong accepted leaf-fault
enumerators in Eq.~\eqref{eq:probpolys}; the last column gives
$g_{15}(F)=\sum_{j=0}^{15}g_jF^j$ in Eq.~\eqref{eq:factorization},
with the row index read as $j$ for that column.
These integers specify the full ideal recurrence, not only its
low-noise expansion.}
\label{tab:coeffs}
\begin{ruledtabular}
\begin{tabular}{rrrr}
$w$ (or $j$)&$c_w$ (correct)&$d_w$ (wrong)&$g_j$\\
\hline
0&1&0&-149949201\\
1&0&0&263381416\\
2&40&0&-2122800736\\
3&32&384&4691993600\\
4&1036&3456&-20373018624\\
5&1984&23808&58392420352\\
6&32824&186624&-238242988032\\
7&121440&1048704&716242124800\\
8&839286&5423232&-2048766509056\\
9&3219072&23625216&4884383924224\\
10&12934488&87471616&-10509663338496\\
11&38209248&272910976&18851022503936\\
12&99569388&692033152&-27066279919616\\
13&195839424&1372460288&28467580108800\\
14&289306056&2025922816&-16854525411328\\
15&283718304&1985266560&8505108987904\\
16&149949201&1049815936&---\\
\end{tabular}
\end{ruledtabular}
\end{table*}

For the three-layer recurrence,
\begin{equation}
 R_3(F)-F=-\frac{(F-1)(8F-1)g_7(F)}{D_3(F)},
 \label{eq:r3factor}
\end{equation}
where $D_3=7^8Z_{\Rthree}>0$ and
\begin{align}
 g_7(F)={}&1458176F^7-884736F^6+548864F^5\nonumber\\
 &-368640F^4+138496F^3-67808F^2\nonumber\\
 &+5440F-6249.
 \label{eq:g7}
\end{align}
We apply Sturm's real-root counting theorem
\cite{basu2006}. The Sturm sequences of $g_7$ and $g_{15}$
have variations
$(4,3)$ and $(8,7)$ at $(0,1)$, respectively. Hence each has exactly
one root in that interval. Exact sign evaluation brackets the
$g_7$ root between $0.61560748080966942$ and
$0.61560748080966943$. For the four-layer map,
\begin{equation}
 \frac{1329775015825}{2199023255552}
 <F_*<
 \frac{664887507913}{1099511627776}.
 \label{eq:rootbracket}
\end{equation}
The two endpoints give opposite signs of $g_{15}$. All decimal
endpoints in the three-layer sign test are interpreted as rationals.

Since $g_{15}(1)=4747561509943>0$ and it has no root above $F_*$,
Eq.~\eqref{eq:factorization} gives $R_4(F)>F$ for $F_*<F<1$.
An iteration started in this interval is increasing and bounded by one.
Its limit is a fixed point by continuity; the root count leaves only
one as a possible limit. The check denominators are positive, so
every finite iteration has positive success probability. This proves
the basin statement used in Sec.~\ref{sec:recurrence}.

The same enumerators provide the local error laws. For $\Rfour$,
$d_0=d_1=d_2=0$ and $d_3=384$. Retaining two additional orders gives
\begin{equation}
 1-R_4(1-\eps)
 =\frac{384}{343}\eps^3+\frac{11520}{2401}\eps^4
 +\frac{218112}{16807}\eps^5+O(\eps^6).
 \label{eq:taylorextended}
\end{equation}
For $\Rthree$, $d_2=48$ and
$1-R_3(1-\eps)=(48/49)\eps^2+(960/343)\eps^3+O(\eps^4)$.
The probability-space results agree exactly with symbolic
reconstruction from Eq.~\eqref{eq:update}.

To make the parity argument of Sec.~\ref{sec:cubic} explicit, the
$\Rthree$ weight-two syndrome counts in integer order $0,\ldots,7$
are
\[
 (20,16,16,0,16,0,0,0),
\]
and its weight-three counts are
\[
 (16,32,32,64,32,64,64,0).
\]
These vectors suffice to recover the low-weight $\Rfour$ table
without enumerating all $8^{16}$ raw syndrome assignments.

\section{Exact evaluation of the initial sequences}\label{app:certificate}

We give the exact calculations for the initial sequences in
Sec.~\ref{sec:certificate}. We first evaluate the balanced benchmark,
then the heterogeneous sequence, keeping track of the output state,
independently prepared inputs, and expected raw-copy cost.

\subsection{Balanced benchmark}

For the word $W_{25}$ in Eq.~\eqref{eq:w25}, the exact output fidelity is
enclosed by
\begin{equation}
 \begin{aligned}
 F_{25}&>0.60550725738718132769784,\\
 F_{25}&<0.60550725738718132769785 .
 \end{aligned}
 \label{eq:certificate}
\end{equation}
The certificate uses enclosing interval arithmetic
\cite{moore2009}, with rational rather than floating-point endpoints.
For an interval $[a,b]$ and positive integer grid denominator $G$,
outward rounding means
\begin{equation}
 [a,b]\longmapsto
 \left[\frac{\lfloor Ga\rfloor}{G},
       \frac{\lceil Gb\rceil}{G}\right].
 \label{eq:rounding}
\end{equation}
Interval multiplication uses the minimum and maximum of the four endpoint
products. Division is used only when the denominator interval excludes
zero, with the analogous four endpoint quotients. The denominators
encountered along the $W_{25}$ trajectory are checked to be strictly
positive.

Starting from the exact isotropic Fourier point at $F=5997/10000$, we
apply Eq.~\eqref{eq:update} with $G=10^{80}$ and round outward after every
coordinate update. Induction over the 25 comparisons therefore encloses
the exact trajectory. The lower endpoint in Eq.~\eqref{eq:certificate} is
then compared by exact cross multiplication with the dyadic upper bound
on $F_*$ from Appendix~\ref{app:polynomials}. The strict inequality
between these two rational numbers is the step that proves basin entry.

We also evaluate the same prefix directly in probability space, without
using the Fourier update. Starting from the eight point probabilities, the
unnormalized accepted output distribution is
\begin{equation}
 \widetilde P(u)=
 \!\!\sum_{\substack{\ip{\ell}{r}=\ip{\ell}{s}\\
 r+s+\ip{\ell}{s}e=u}} P(r)P(s),\qquad
 P'(u)=\frac{\widetilde P(u)}{\sum_t\widetilde P(t)} .
 \label{eq:directinterval}
\end{equation}
The calculation uses rational intervals with outward rounding on a
$10^{-100}$ grid. Its final enclosure is compatible with
Eq.~\eqref{eq:certificate} and independently satisfies the basin-entry
inequality. This second calculation is useful because it does not call
the Fourier-coordinate update used by the first implementation.

For comparison, the shorter sequence mentioned in the main text is
\begin{equation}
 W_{14}=(6,1,5,1,2,7,4,5,2,1,6,5,2,1).
 \label{eq:w14}
\end{equation}
At $F=3/5$ direct rational arithmetic gives
\begin{equation}
 W_{14}(\pi_{3/5})(0)
 =0.6091260418296702\ldots>0.609>F_* .
 \label{eq:w14exact}
\end{equation}
The reduced numerator and denominator each have 72648 bits. Serializing
their minimal unsigned big-endian representations, separated by the ASCII
byte \texttt{/}, gives the SHA-256 digest
\begin{center}
\footnotesize\ttfamily
6d203367282e4715b11351dd959dbae87ca\\
967b5603bbc6dc82a10d2f596fa38
\end{center}
with the line break removed. The digest is included only as a regression
identifier; the mathematical statement is the exact rational inequality
in Eq.~\eqref{eq:w14exact}.

The product of conditional acceptance probabilities is accumulated with
the same outward arithmetic. At the $W_{25}$ endpoint, a short enclosing
display of the pool yield is
\begin{equation}
 \begin{aligned}
 Y(W_{25})&>0.00000000000005684665,\\
 Y(W_{25})&<0.00000000000005684666 .
 \end{aligned}
 \label{eq:yieldinterval}
\end{equation}
The upper endpoint is rounded upward when printed. These numbers underlie
the resource estimate in Sec.~\ref{sec:cost} but are not needed to prove
distillability.

The verification code uses Python integer arithmetic,
\texttt{fractions.Fraction}, and exact SymPy expressions
\cite{meurer2017}. The operation-noise continuation additionally uses
\texttt{mpmath} \cite{mpmath2023}. The reference checks were run with Python 3.13.5,
SymPy 1.14.0, and mpmath 1.3.0. Scientific verification conditions raise
explicit exceptions and remain active under Python's optimized
\texttt{-O} mode. Floating-point or high-precision root-finding results
are not used in the endpoint proof. The independent probability-space
implementation also checks the operator-sign covariance and the
low-weight fault enumerators used elsewhere in the paper.

\subsection{Heterogeneous endpoint and independent calculation}

Fix $F=a/b$ in lowest terms. Use the raw integer weights
\begin{equation}
 w^{(0)}=(7a,b-a,b-a,b-a,b-a,b-a,b-a,b-a),
 \label{eq:rawinteger}
\end{equation}
whose sum is $7b$. At stage $j$, use $v=w^{(j-1)}$ for
$\mathsf S$ and $v=w^{(0)}$ for $\mathsf R$, and compute
\begin{equation}
 w_u^{(j)}
 =\!\!\sum_{\substack{\ip{\ell_j}{r}=\ip{\ell_j}{s}\\
 C_{\ell_j,e(\ell_j)}(r,s)=u}}
 w_r^{(j-1)}v_s .
 \label{eq:integerupdate}
\end{equation}
No normalization is performed in this recursion. Each product counts
independent assignments on the two child subtrees. Since all raw weights
are positive at the endpoints used here, the all-zero assignment gives
positive accepted weight at every stage. The exact normalized state and
the conditional acceptance probability are
\begin{equation}
 \begin{aligned}
 P_j(u)&=\frac{w_u^{(j)}}{\sum_t w_t^{(j)}},\\
 A_j&=\frac{\sum_t w_t^{(j)}}
 {(\sum_r w_r^{(j-1)})(\sum_s v_s)} .
 \end{aligned}
 \label{eq:integernormalized}
\end{equation}
Induction on the tree shows that these are the same state and acceptance
as the normalized syndrome update. Separately prepared subtrees use
disjoint raw blocks. Thus no copying of intermediate quantum states is
assumed.

At $F=5995/10000$, the exact rational output has a $25520$-bit
numerator and a $25521$-bit denominator. Directed integer division gives
\begin{equation}
 \begin{aligned}
 \mathcal H_{12}(\pi_F)(0)&>0.605026344233701236200599895568,\\
 \mathcal H_{12}(\pi_F)(0)&<0.605026344233701236200599895569 .
 \end{aligned}
 \label{eq:h12certificate}
\end{equation}
The first bound is greater than the dyadic upper bound in
Eq.~\eqref{eq:rootbracket}, as checked by integer cross multiplication.
The independent recheck compares all eight probabilities with the
Fourier implementation at every one of the twelve stages, including
the nonidentical-input stage. It also reproduces the final fraction
from direct normalized probability propagation.

The regression digest of the final reduced fraction, using the byte
serialization specified above, is
\begin{center}
\footnotesize\ttfamily
2ccfebe48dd4c020d42e3991fb9e354c9e9a\\
1e692f17b0bda28aed044ca43a51
\end{center}
with the line break removed. Table~\ref{tab:h12stages} records the
intermediate trajectory. The digest identifies an exact result but is
not its mathematical proof.

At the common comparison fidelity $F=5997/10000$,
\begin{equation}
 \begin{aligned}
 \mathcal H_{12}(\pi_F)(0)&>0.611888396112547544970046714887,\\
 \mathcal H_{12}(\pi_F)(0)&<0.611888396112547544970046714888,\\
 Y(\mathcal H_{12})&>0.0000009645966666931788100,\\
 Y(\mathcal H_{12})&<0.0000009645966666931788101 .
 \end{aligned}
 \label{eq:h12common}
\end{equation}
Combining these with Eq.~\eqref{eq:yieldinterval} gives a yield-ratio
enclosure from $16968396.501$ to $16968399.487$.

\subsection{One-stage self-pairing control}
\label{app:control}

Define $\mathcal H_{12}^{\rm bal}$ by changing only the fourth step
of Eq.~\eqref{eq:h12} from $\mathsf R_7$ to $\mathsf S_7$.
The integer recursion \eqref{eq:integerupdate}, with $v=w^{(j-1)}$
at every stage, gives at $F=5995/10000$
\begin{equation}
\begin{aligned}
 \mathcal H_{12}^{\rm bal}(\pi_F)(0)
   &=0.551387370616520\ldots,\\
 \mathcal H_{12}(\pi_F)(0)
   &=0.605026344233701\ldots .
\end{aligned}
\label{eq:controlappendix}
\end{equation}
At $F=5997/10000$ the corresponding fidelities are
$0.562269479262795\ldots$ and $0.611888396112548\ldots$.
Every number is obtained from integer-weight propagation and hence
represents a rational value before decimal rounding. The code archive
contains an independent verifier of this replacement, including the
expected retry costs. No optimization over balanced words is involved.

\subsection{Retry cost and finite-tree interpretation}

The cost formula does not require the cost of one attempt to be
independent of its own success. Let $K_i$ be the cost of attempt $i$ and
$T$ the index of the first successful attempt. Independent fresh attempts
have success probability $A$ and mean attempt cost $k$.
The event $T\ge i$ depends only on preceding attempts and is independent
of $K_i$. Therefore, summing nonnegative costs,
\begin{equation}
 \begin{split}
 \mathbb E\sum_{i=1}^{T}K_i
 &=\sum_{i\ge1}\mathbb E[K_i\mathbf1_{\{T\ge i\}}]\\
 &=k\sum_{i\ge1}(1-A)^{i-1}=\frac{k}{A}.
 \end{split}
 \label{eq:retryproof}
\end{equation}
Taking $k=2C_{j-1}$ or $k=C_{j-1}+1$ gives
Eq.~\eqref{eq:heterocost}. The returned syndrome distribution is
unchanged by retry: if one attempt accepts with syndrome $u$ with
probability $B(u)$, its probability after retry is
$\sum_{i\ge1}(1-A)^{i-1}B(u)=B(u)/A$.

For $\mathcal H_{12}$, the cost also has the explicit form
\begin{equation}
 C(\mathcal H_{12})
 =\frac{2048/(A_1A_2A_3)+256}{A_4A_5\cdots A_{12}}.
 \label{eq:h12closedcost}
\end{equation}
The first three self-pairings cost $8/(A_1A_2A_3)$. The raw-input
stage adds one and divides by $A_4$; the remaining eight stages
multiply by $2^8/(A_5\cdots A_{12})$. At $F=0.5995$, exact evaluation
gives
\begin{equation}
 \begin{aligned}
 C&>1051931.361812434,\\
 C&<1051931.361812435,\\
 Y&>0.00000095063236661852,\\
 Y&<0.00000095063236661853 .
 \end{aligned}
 \label{eq:h12costenclosure}
\end{equation}
Reverse propagation of the required stream sizes gives the same cost,
independently of the forward cost recursion.

For a preassigned tree the success probabilities instead obey
$Z_0=1$, $Z_j=A_jZ_{j-1}^2$ at an $\mathsf S$ stage and
$Z_j=A_jZ_{j-1}$ at an $\mathsf R$ stage. They are positive.
A binary tree with $N$ raw leaves has $N-1$ comparison nodes, giving
$2303$ nodes for this construction. Conditional states agree with
recursive retry, but the attempt probabilities and raw costs do not.
The finite-protocol distillability statement can use this fixed tree
and any finite number of subsequent recurrence rounds. It does not
rely on an unbounded retry time.

\begin{table}[tbp]
\caption{Stage-by-stage heterogeneous certificate at raw fidelity
$F=0.5995$. $F_j$ and $A_j$ are displayed with rounding; the verifier
uses exact rationals. $N_j$ counts the raw leaves of a fully accepted
stage-$j$ construction.}
\label{tab:h12stages}
\setlength{\tabcolsep}{4pt}
\begin{ruledtabular}
\begin{tabular}{ccccc}
$j$&Step&$F_j$&$A_j$&$N_j$\\
\hline
1&$\mathsf S_1$&$0.570633104$&$0.647036898$&2\\
2&$\mathsf S_2$&$0.563467337$&$0.603281982$&4\\
3&$\mathsf S_4$&$0.576392932$&$0.558619001$&8\\
4&$\mathsf R_7$&$0.582290290$&$0.604935566$&9\\
5&$\mathsf S_1$&$0.569560078$&$0.616019516$&18\\
6&$\mathsf S_3$&$0.578260519$&$0.570941954$&36\\
7&$\mathsf S_7$&$0.586642668$&$0.578509631$&72\\
8&$\mathsf S_6$&$0.580974805$&$0.605089262$&144\\
9&$\mathsf S_3$&$0.586088605$&$0.586715943$&288\\
10&$\mathsf S_4$&$0.589206912$&$0.593676622$&576\\
11&$\mathsf S_1$&$0.596700723$&$0.591390727$&1152\\
12&$\mathsf S_6$&$0.605026344$&$0.597831150$&2304\\
\end{tabular}
\end{ruledtabular}
\end{table}

\section{The convex-Gaussian boundary and the no-go statement}
\label{app:nogo}

For a four-mode even-sector state $\rho$, we use the criterion of
Oszmaniec, Gutt, and Ku\'s \cite{oszmaniec2014}, Eqs.~(4) and (5),
based on conjugated-state concurrence \cite{uhlmann2000}. It uses the ordered nonnegative eigenvalues
$\lambda_1\geq\cdots\geq\lambda_8$ of
$\sqrt{\rho\widetilde\rho}$:
\begin{equation}
 C_+(\rho)=\max\{0,\lambda_1-\lambda_2-\cdots-\lambda_8\}.
 \label{eq:criterion}
\end{equation}
The state is convex-Gaussian if and only if this quantity vanishes.

The conjugation changes the explicit factors of $i$ in the Majorana
expansion. The chosen $S_j$ and $Q$ have real coefficients and
degrees four and eight. Therefore every $\Pi_s$ is invariant and
$\widetilde{\sigma_F}=\sigma_F$. Positivity gives
$\sqrt{\sigma_F\widetilde{\sigma_F}}=\sigma_F$, with eigenvalues
$F$ and $q=(1-F)/7$ repeated seven times. Their sum is one.
Equation~\eqref{eq:concurrence} follows. Since $q\leq1/7<1/2$,
the criterion vanishes exactly for $F\leq1/2$.

It remains to connect membership to a distillation obstruction.
A tensor product of convex-Gaussian states is a mixture of pure
Gaussian product states. A fixed branch of an FLO protocol maps
each Gaussian component to a nonnegative multiple of a Gaussian
state, possibly mixed when modes are discarded
\cite{bravyi2005,demelo2013}. A mixed Gaussian state itself has a
convex decomposition into pure Gaussian states, as seen by
diagonalizing its quadratic normal form. Normalizing a nonzero
branch therefore leaves a convex-Gaussian output. This argument
does not assume that a partial trace preserves purity.

For the target projector, $C_+(\Pi_0)=1$. The four-mode squared-fidelity
formula, Eq.~(6) of Ref.~\cite{oszmaniec2014}, then gives
\begin{equation}
 \max_{\tau\in\cG}\bra{a_8}\tau\ket{a_8}=\frac12 .
 \label{eq:gaussianoverlap}
\end{equation}
An odd-parity component has zero overlap with the even target and
cannot increase this maximum. Hence every conditional four-mode
output from a convex-Gaussian input has fidelity at most $1/2$.
This proves the no-go side of Eq.~\eqref{eq:regions}.

\section{Operation-noise calculations and stochastic circuit faults}
\label{app:noise}

\subsection{Effective syndrome channel and numerical continuation}

For the channel in Eq.~\eqref{eq:noise}, convolution with
$\nu_\lambda$ multiplies every nontrivial syndrome correlator by the
same factor. If $y_v$ denotes the ideal output of
Eq.~\eqref{eq:update}, then
\begin{equation}
 x'_0=1,\qquad
 x'_v=\kappa_\lambda y_v\quad(v\ne0),\qquad
 \kappa_\lambda=1-\frac87\lambda .
 \label{eq:noisyupdate}
\end{equation}
The acceptance probability of the comparison is unchanged at that node,
although all later comparisons act on the noisy output. Applying
Eq.~\eqref{eq:noisyupdate} after every successful comparison of the word
$(1,2,4,7)$ and isotropizing at the end defines $R_{4,\lambda}$.

High-precision root continuation gives the fixed points listed in
Table~\ref{tab:noisyroots}. The derivative of the recurrence identifies
the lower branch as unstable and the upper branch as attracting. These
values are numerical; only the noiseless fixed point used in the ideal
threshold proof is established by the exact polynomial analysis of
Appendix~\ref{app:polynomials}.

\begin{table}[!tbp]
\caption{Numerical fixed points of the effective-noise recurrence.}
\label{tab:noisyroots}
\begin{ruledtabular}
\begin{tabular}{ccc}
$\lambda$&$F_-(\lambda)$&$F_+(\lambda)$\\
\hline
$0$&$0.604711665721532$&$1$\\
$10^{-4}$&$0.604885431446267$&$0.999757042328097$\\
$10^{-3}$&$0.606459520811497$&$0.997561265472276$\\
$5\,10^{-3}$&$0.613690033554464$&$0.987589998096975$\\
$10^{-2}$&$0.623332228826214$&$0.974571486065393$\\
$2\,10^{-2}$&$0.645211515251389$&$0.946095042634382$
\end{tabular}
\end{ruledtabular}
\end{table}

For the upper branch, put
$h(F,\lambda)=R_{4,\lambda}(F)-F$. The map is analytic near
$(F,\lambda)=(1,0)$, with $h(1,0)=0$ and
$\partial_F h(1,0)=-1$. The implicit-function theorem
\cite{krantz2013} gives a locally unique smooth fixed-point branch.
Its derivative is
$F_+'(0)=-\partial_\lambda h/\partial_F h=-17/7$.
Moreover $\partial_F R_{4,0}(1)=0$, so continuity gives an attracting
branch for sufficiently small positive $\lambda$.
This yields Eq.~\eqref{eq:floor} without using numerical continuation
as an existence proof.

Implicit differentiation at the noiseless unstable root gives the
coefficient quoted in Eq.~\eqref{eq:lowerbranch}; its displayed decimal
value is evaluated numerically.
Solving the fixed-point and tangency equations simultaneously gives
\begin{equation}
 \begin{aligned}
 \lambda_c&=0.0474853011598231\ldots,\\
 F_c&=0.787046431803765\ldots .
 \end{aligned}
 \label{eq:fold}
\end{equation}
This is the numerical fold observed in the effective recurrence and is
not used as an exact global classification of noisy trajectories.

For the finite prefixes, we apply the same noisy update after every
comparison and solve for the point where the prefix output equals the
lower fixed-point branch. The two values used in the main text are
\begin{align}
 \lambda_{\rm cross}(W_{25},0.5997)
 &=1.00702159328777\ldots\times10^{-7},\nonumber\\
 \lambda_{\rm cross}(W_{14},0.6)
 &=2.51686321131286\ldots\times10^{-5}.
 \label{eq:prefixcrossings}
\end{align}
For the heterogeneous schedule, the raw block at the fourth stage
is still an independent copy of $\pi_F$. It does not acquire an extra
operation fault before use; the channel is applied to the retained
output of the comparison. All processed inputs carry the faults from
their own histories. The corresponding local crossings are
\begin{equation}
 \begin{aligned}
 \lambda_{\rm cross}(\mathcal H_{12},0.5997)
 &=1.2931807295139568\ldots\times10^{-4},\\
 \lambda_{\rm cross}(\mathcal H_{12},0.5995)
 &=5.65511716055258\ldots\times10^{-6}.
 \end{aligned}
 \label{eq:heterocrossings}
\end{equation}
The same fixed schedule has a numerical noiseless basin crossing near
$F=0.59949086085192491018$. This is not used in the certificate
\eqref{eq:h12certificate}.

These roots are independently checked in probability space by applying
$P(u)\mapsto(1-\lambda)P(u)+(\lambda/7)[1-P(u)]$ after each accepted
comparison. The fixed-point and prefix-crossing equations are solved
jointly at 80-digit precision and repeated at 110 digits. Residuals
and the unstable-branch derivative are checked, as are the positive
and negative margins at 90 and 110 percent of each crossing.
This verifies local numerical behavior, not global uniqueness of all
crossings or an interval-certified noisy basin.

\subsection{Comparison of single operation faults with the three-layer round}

The same first-order calculation can be performed for $\Rthree$.
After a single inserted fault $d\ne0$, each later check pairs the
corrupted block with a perfect one. It accepts exactly when its checked
parity of $d$ vanishes, and then leaves $d$ unchanged.
If the remaining check labels have binary rank $r$, their common kernel
has $2^{3-r}$ elements, one of which is zero.
For $(1,2,4)$ the later ranks are $2,1,0$, and the numbers of fault
locations are $4,2,1$. The harmful accepted count is therefore
$4(2^1-1)+2(2^2-1)+(2^3-1)=17$.
Of the $49$ labeled single faults, the remaining $32$ reject.
For $(1,2,4,7)$ the later ranks are $3,2,1,0$; the corresponding
count is $8(2^0-1)+4(2^1-1)+2(2^2-1)+(2^3-1)=17$.
Thus
\begin{align}
 Z_{3,\lambda}(1)&=1-\frac{32}{7}\lambda+O(\lambda^2),\nonumber\\
 1-R_{3,\lambda}(1)&=\frac{17}{7}\lambda+O(\lambda^2).
 \label{eq:r3operationfloor}
\end{align}
A labeled single fault has probability $\lambda/7+O(\lambda^2)$;
two or more faults contribute only at second order. Normalization
does not change the linear term because acceptance tends to one.
The three- and four-layer rounds have the same leading output-error
coefficient in this effective model. This does not imply equal
acceptance penalties, equal noisy basins, or equal hardware thresholds.

\subsection{The terminal channel and common-mode shifts}

The terminal-channel limitation follows without a root calculation.
If $f$ is the target population just before the final unheralded
channel, then
\begin{equation}
 F_{\rm final}
 =\left(1-\frac87\lambda\right)f+\frac{\lambda}{7}
 \leq1-\lambda,\qquad 0\leq\lambda\leq7/8.
 \label{eq:terminalfloor}
\end{equation}
The inequality uses $f\leq1$ and the nonnegative coefficient of $f$.
A final isotropization preserves the fidelity. A different circuit
with additional fault information is not covered by this argument.

A common-mode syndrome shift illustrates a limitation of any
independent-error interpretation. For every admissible check,
\begin{equation}
 C_{\ell,e}(r+d,s+d)
 =C_{\ell,e}(r,s)+\ip{\ell}{d}e .
 \label{eq:common}
\end{equation}
Hence a shared shift does not change the acceptance condition. It only
changes the retained output frame when $\ip{\ell}{d}=1$.

\subsection{Conditioning on faulty accepted branches}

For a fixed input, let
$\Tr\mathcal E_{\rm acc}^{\rm id}(\rho)=Z$ and
$\Tr[\Pi_0\mathcal E_{\rm acc}^{\rm id}(\rho)]=ZR$.
Under Eq.~\eqref{eq:mixture}, write
$b=\Tr\mathcal E_{\rm acc}^{\rm bad}(\rho)\leq1$ and
$g=\Tr[\Pi_0\mathcal E_{\rm acc}^{\rm bad}(\rho)]\geq0$.
The actual conditional fidelity is
\begin{equation}
 \frac{(1-t)ZR+tg}{(1-t)Z+tb}
 \geq\frac{(1-t)ZR}{(1-t)Z+t}.
 \label{eq:mixtureproof}
\end{equation}
The right-hand side decreases with $t$, so
$t\leq\eta_{\rm tree}$ proves Eq.~\eqref{eq:badbound}.
Solving for the condition that this lower bound exceed the input
fidelity $F$ gives
\begin{equation}
 \eta_{\rm tree}<
 \frac{Z[R-F]}{F+Z[R-F]}.
 \label{eq:etabound}
\end{equation}
This is the expression used for the numerical values in the main
text. The fifteen locations are effective comparison nodes, not a
count of all elementary exchanges and readouts in a routed device.

The ideal-branch assumption is essential. An input-dependent fault
flag can preferentially remove correct accepted events. For
example, in an $\Rfour$ tree at $F=3/4$, flag only a subset of
all-zero leaf assignments, of total probability $\eta=10^{-6}$,
and replace their retained output by a wrong syndrome without
changing acceptance. This is possible because $F^{16}>\eta$.
The actual fidelity is then
\[
 F_{\rm actual}=\frac{Z_4R_4-\eta}{Z_4}.
\]
Exact rational evaluation gives approximately $0.9417165673$,
whereas the right-hand side of Eq.~\eqref{eq:badbound} would be
$0.9417215436$. A marginal bound on the fault probability is
satisfied, but the asserted lower bound would fail. Such selection
is excluded by Eq.~\eqref{eq:mixture}. Correlated locations and
state-dependent selection of faulty branches are different
assumptions and must not be conflated.

\section{Isotropization after every comparison}
\label{app:onecheck}

For comparison with the multi-layer schedules, consider the more
restrictive rule of isotropizing after each individual comparison.
This is not Bravyi's original recurrence, which randomizes only after
a complete round \cite{bravyi2006}.
The effect can also be seen in a general syndrome dimension. Consider an isotropic distribution on
$\Ftwo^k$, $k\geq2$, with nonzero Fourier coefficients equal to $x$.
After one identical-input comparison, the checked coefficient is
$2x/(1+x^2)$ and the remaining $2^k-2$ nontrivial coefficients are
$2x^2/(1+x^2)$. Immediate averaging over $\GL(k,2)$ therefore gives
\begin{equation}
 T_k(x)=\frac{2x+2(2^k-2)x^2}{(2^k-1)(1+x^2)}.
 \label{eq:onecheck}
\end{equation}
Its fixed points in $[0,1]$ are
\begin{equation}
 x=0,\qquad x=\frac{2^k-3}{2^k-1},\qquad x=1.
 \label{eq:onecheckfixed}
\end{equation}
For $k=3$, the nontrivial boundary is $x=5/7$, corresponding to
$F=3/4$. Isotropizing after every comparison therefore loses the
unequal syndrome populations used by the multi-layer sequences.

\section{Numerical comparison of complete purification schedules}
\label{app:factorycomparison}

For each round $W$, let $R_W(f)$ be the conditional output fidelity
from the isotropic input of fidelity $f$, and let $c_W(f)$ be the
expected number of such inputs consumed under elementary-comparison
retry. Start at $f_0=F$ and $C_0=1$ when no prefix is used.
For the heterogeneous construction, initialize instead with its
calculated seed fidelity and expected raw cost.
After each complete recurrent round, set
\begin{equation}
 f_{k+1}=R_W(f_k),\qquad
 C_{k+1}=C_k\,c_W(f_k).
 \label{eq:factorycostupdate}
\end{equation}
Each input of the new round is produced independently with mean raw
cost $C_k$, so the node-by-node retry proof in
Appendix~\ref{app:certificate} gives the second identity.
The output is isotropized after the prefix and after each complete round.

We use the first complete round for which $1-f_k\le10^{-6}$.
Table~\ref{tab:matchedcost} gives examples where both recurrent
constructions work. At the two heterogeneous endpoint examples the
same prescribed procedure gives
\begin{equation}
\begin{array}{c|c|c|c}
F& n & C &1-f_n\\ \hline
0.5995&8&2.77516\,10^{20}&6.87828\,10^{-13}\\
0.5997&5&4.52768\,10^{14}&2.43181\,10^{-7}
\end{array}
\label{eq:endpointfactorycost}
\end{equation}
Here $n$ excludes the prefix. These large costs include subsequent
purification of the seed; they must not be confused with the
roughly million-copy basin-entry cost.

The calculations are performed independently in probability space
using Eq.~\eqref{eq:check} and in Fourier coordinates using
Eq.~\eqref{eq:update}. They are repeated at 80 and 110 digits.
The two implementations agree on the stopping round, the output
fidelity, and the expected raw-copy cost to the displayed precision.
These are numerical comparisons, not exact rational enclosures.
They do not optimize schedules, pooling at finite batch size,
memory, routing, or microscopic operation noise.

\subsection{Asymptotic accuracy cost at fixed seed fidelity}

For completeness, write $e_n=1-f_n$ after $n$ four-layer rounds.
The expansion in Eq.~\eqref{eq:taylorextended} supplies constants
$a,b>0$ and a sufficiently small neighborhood of zero in which
$a e_n^3\leq e_{n+1}\leq b e_n^3$.
Once a fixed seed reaches this neighborhood, put $L_n=-\log e_n$.
Then
\[
 3L_n-\log b\leq L_{n+1}\leq3L_n-\log a.
\]
Choose the starting index sufficiently late that
$L_0>\max(0,\tfrac12\log b)$.
Summing these inequalities gives positive constants $d_1,d_2$
with $d_1 3^n\leq L_n\leq d_2 3^n$ for all sufficiently large $n$.
Consequently the first $n$ with $e_n\leq\delta$ satisfies
$n=\log_3\log(1/\delta)+O(1)$.

Let $\alpha(f)$ be the product of the four conditional layer
acceptances of one round. Analyticity and $\alpha(1)=1$ give
$0\leq1-\alpha(1-e)\leq K e$ near zero.
Shrinking that neighborhood so that $b e^2\leq1/2$ shows
$e_{n+1}\leq e_n/2$, hence $\sum_n e_n<\infty$.
For all sufficiently late rounds, $1-\alpha(f_n)\leq1/2$ and
$-\log\alpha(f_n)\leq2[1-\alpha(f_n)]$.
Thus the product of the acceptances over those rounds has a
strictly positive limit. Earlier rounds contribute only a fixed
positive factor. The expected input cost is therefore bounded by
a constant times $16^n$, proving Eq.~\eqref{eq:overhead}.
All constants depend on the fixed seed; the result is not uniform
as the seed approaches $F_*$.

\bibliography{majorana_refs}
\end{document}